\documentclass[aps,prc,preprint,showpacs]{revtex4-1}

\usepackage[usenames,dvipsnames]{xcolor}
\usepackage{graphicx}
\usepackage{amssymb}
\usepackage{amsmath} 
\usepackage{mathtools}
\usepackage[T1]{fontenc} 
\usepackage[utf8]{inputenc}

\DeclarePairedDelimiter\abs{\lvert}{\rvert}

\newcommand{\coll}[2]{#1$\,+\,$#2}              
\newcommand{\collpt}[4]{$^{#1}$#2$\,+^{#3}$#4}  

\newcommand{\imnncs}{\ensuremath{\sigma^\textrm{med}_\textsc{NN}}}

\begin{document}
\title{Shear viscosity from nuclear stopping}
\author{Brent Barker}
\affiliation{Department of Biological, Chemical, \& Physical Sciences, Roosevelt University, Chicago, IL 60605, USA}
\email{bbarker@roosevelt.edu}
\author{Pawel Danielewicz}
\affiliation{National Superconducting Cyclotron Laboratory and Department of Physics \& Astronomy, Michigan State University, East Lansing, MI 48824, USA}
\date{\today}

\begin{abstract}

Within a Boltzmann transport model, we demonstrate correlation between stopping observables and shear viscosity in central nuclear collisions at intermediate energies (on the order of 10--1000 MeV/nucleon).
The correlation allows us to assess the viscosity of nuclear matter, by tuning the in-medium nucleon-nucleon cross section in our transport model to agree with nuclear stopping data. We also calculate the ratio of shear viscosity to entropy density to determine how close the system is to the universal quantum lower limit proposed in the context of ultrarelativistic heavy ion collisions.

\end{abstract}

\maketitle

\section{Introduction}

Collisions between atomic nuclei at intermediate energies are often used to infer the bulk properties of nuclear matter.
Central among the bulk properties is the nuclear equation of state (EOS), which pertains to the state of stationary equilibrium of the matter and contains no information regarding the pace at which the equilibrium is reached.
Collective flow observables were successfully exploited to infer the EOS.
On the other hand, bulk properties that are tied to equilibration rate include transport coefficients such as shear viscosity and heat conduction.
In particular, shear viscosity is tied to momentum transport in a medium and, among reaction observables, it is natural to link it to stopping observables that quantify dissipation of momentum.

Knowledge of the shear viscosity is important for understanding the evolution of supernovae, the stability of rotating neutron stars, and the formation of black holes.
Besides its immediate practical importance, there have been conjectures regarding a fundamental quantum lower limit on the ratio of shear viscosity to entropy density ($\eta / s$) in a wide range of media \cite{danielewicz_dissipative_1985,kovtun_viscosity_2005, schafer_fluid_2014}. Among other situations, the limit is thought to be approached in the quark-gluon plasma and accessed in ultrarelativistic heavy ion collisions \cite{lacey_has_2007}.
The question remains as to whether freeing quark degrees of freedom is needed to approach such a limit in collisions.
We will make a quantitative assessment of how close nuclear matter, as seen in these lower-energy collisions, is to this ``perfect liquid'' limit.

In the present work, we use stopping, i.e.\ the degradation of the projectile longitudinal momentum due to interaction with the target, to constrain the elastic part of the in-medium nucleon-nucleon cross section, \imnncs{}, in a \textsc{BUU} transport model.
These constraints are not without ambiguity. Different observables might lead to different conclusions. Therefore, we consider different stopping observables for different systems at different energies. What's more, different strategies for modifying the cross section in the nuclear medium can lead to the same degree of stopping, unsurprisingly.
Consequently, we inspect whether shear viscosity is actually correlated with the stopping observables in collisions.
We calculate the viscosity in a manner consistent with the Boltzmann equation used to describe the collisions and find a strong correlation between the predicted stopping observables and the magnitude of the predicted shear viscosity coefficients.
The correlation suggests a robustness in the conclusions on the viscosity, even when cross sections are not easy to pin down unambiguously based on the data alone.

\section{Boltzmann-Uehling-Uhlenbeck equation}\label{sec:buu}

To model central nuclear reactions and predict observables, we use a set of Boltzmann-Uehling-Uhlenbeck (\textsc{BUU}) equations, one for each species X, describing the time evolution of
a Wigner quasi-probability distribution in phase space,
$f_\mathrm{X} \equiv f_\mathrm{X}(\vec{r},\vec{p},t)$:
\begin{equation}\label{eqn:buu}
\frac{\partial f_\mathrm{X}}{\partial t}
+ \frac{\partial \epsilon_{\vec{p}}}{\partial \vec{p}} \frac{\partial f_\mathrm{X}}{\partial \vec{r}}
- \frac{\partial \epsilon_{\vec{p}}}{\partial \vec{r}} \frac{\partial f_\mathrm{X}}{\partial \vec{p}}
= I_{\mathrm{X}, \textrm{elastic}} + I_{\mathrm{X}, \textrm{inelastic}} \,.
\end{equation}
A prototype equation for the above is the Vlasov equation (single-particle Liouville equation), with vanishing r.h.s., describing the single-particle evolution of a phase space density in a mean field.
In the above, $\dfrac{\partial \epsilon_{\vec{p}}}{\partial \vec{p}}$ is the single-particle velocity, and $\dfrac{\partial \epsilon_{\vec{p}}}{\partial \vec{r}}$ is the force due to the mean field.

The r.h.s.\ of Eq.~\ref{eqn:buu} takes into account the effects of elastic and inelastic collisions. The elastic contribution can be expressed as
\begin{equation}\label{eq:buu-elastic}
 I_{\mathrm{X}, \textrm{elastic}} = \sum_Y \frac{g_\mathrm{X}}{(2\pi\hbar)^3} \int \mathrm{d}\vec{p}_Y \,\mathrm{d}\Omega \,\, v_{XY} \,\, \frac{\mathrm{d} \sigma}{\mathrm{d} \Omega}
  \left( \tilde{f}_{\mathrm{X}}^{} \tilde{f}_\mathrm{Y}^{} f_\mathrm{X}' f_\mathrm{Y}' - \tilde{f}_\mathrm{X}' \tilde{f}_\mathrm{Y}' f_\mathrm{X}^{} f_\mathrm{Y}^{} \right) \, .
\end{equation}
The first term accounts for particles with momenta $\vec{p}\phantom{'}'_\mathrm{X}$ and $\vec{p}\phantom{'}'_\mathrm{Y}$ colliding and acquiring the final momenta $\vec{p}_\mathrm{X}$ and $\vec{p}_\mathrm{Y}$, thus increasing the occupancy $f_\mathrm{X}$ (gain).
The second term describes, correspondingly, a decrease in the occupancy $f_\mathrm{X}$ in a reverse process (loss).
Here, for nucleons, $\tilde{f}_\mathrm{X} \equiv 1-f_\mathrm{X}$ represents the Pauli principle blocking scattering into the final state $\vec{p}_\mathrm{X}$.
The rate of scattering is governed by the elastic NN cross section $\dfrac{\mathrm{d} \sigma}{\mathrm{d} \Omega}$ (here, a function of relative momentum and the scattering angle $\theta$; $\Omega \equiv (\theta, \varphi)$).
It is this cross section of which modifications by in-medium effects are explored in Section~\ref{sec:cs-reductions}.

The second, inelastic term on the r.h.s.\ of Eq.~\ref{eqn:buu} represents interactions that create or annihilate particles of the given species.
There has been some work in producing medium modifications to inelastic processes \cite{li_isospin_2017}.
However, in the following sections, we only consider modification of elastic cross sections. Therefore, we must limit drawing conclusions to regimes of reaction dynamics where inelastic processes do not significantly affect the dynamics. Once beam energies are high enough, for example, pions are produced early in the collision. This affects the stopping, so until inelastic cross sections are also addressed, we restrict ourselves to lower energies. Formation and breakup of nuclear clusters is an inelastic process too, but we restrict this process to low densities, so that it plays a role only after the dynamics significant to stopping have taken place.

An implementation of a time-dependent solution to the Boltzmann equation set by Danielewicz and collaborators \cite{danielewicz_production_1991, danielewicz_blast_1992, pan_sideward_1993,danielewicz_determination_2000, danielewicz_hadronic_2002, barker_dissipation_2014}, often termed pBUU, is used to describe nuclear collisions.
In this implementation, the Wigner distributions are represented by a large number of test particles.
These particles move along classical trajectories under the influence of the mean field and then encounter binary collisions on a statistical basis with other test particles that are close to them in position space.
With an increase of test particle number, the simulation converges on a better sampled, stable solution \cite{bertsch_guide_1988}.

The single-particle energies $\epsilon_{\vec{p}}$ in Eq.~\ref{eq:buu-elastic} are derived from an assumed energy functional $\mathcal{E}\{f\}$ \cite{danielewicz_determination_2000} that accounts for modifications of the particle energies from free-space values $\epsilon^0_{\vec{p}}$ due to the average effect of interactions with particles in the medium.
The mean-field potential is $U=\epsilon_p - \epsilon_p^0$.
Unless otherwise indicated, we employ in the calculations an energy functional that yields a soft equation of state (EOS) and momentum-dependent $U$.
In the literature, the abbreviation ``SM'' is attributed to such functionals.

\subsection{Impact parameter selection}

Throughout this work, we will be comparing our simulation results to experimental data. In experiment, a range of impact parameters is selected for analysis. Most often, it is uncertain what precisely the distribution of those impact parameters is. In any single transport simulation, the initial state is prepared with one specific
impact parameter. To save computation time, an effective impact parameter, $b_\mathrm{eff}$, is commonly chosen that represents the median in probability for the impact parameter range. For a range bounded by $b_\mathrm{min}$ and $b_\mathrm{max}$, the effective impact parameter $b_\mathrm{eff}$ is normally taken from
\begin{equation}
 \begin{split}
  \pi b_\mathrm{eff}^2 &= \frac{\pi b_\mathrm{min}^2 + \pi b_\mathrm{max}^2}{2} \,, \\ \\
      b_\mathrm{eff} &= \sqrt{\frac{1}{2} \left( b_\mathrm{min}^2 + b_\mathrm{max}^2 \right)} \,.
 \end{split}
\end{equation}

In studies of the central collisions, often experimental ranges effectively start at $b_\mathrm{min}=0$, so $b_\mathrm{eff} = b_\mathrm{max}/\sqrt{2}$. We have tested in several cases that such a single parameter can indeed adequately represent the range, in that results from one parameter agree to a satisfactory degree with those from combining calculations from impact parameters spanning the range.

\section{Shear viscosity}

An elementary setting for introducing the concept of viscosity is that of laminar shear in a macroscopic system.
Consider two plates, with a medium between them, moving in antiparallel directions, in the steady state.
The layer adjacent to one plate induces a shear stress, $\tau$, on the layer below it, causing that layer to have a velocity $v(y-\mathrm{d}y) < v(y)$.
That layer induces a shear stress on the layer under it, and so on.
In the linear response approximation, these velocities can be related using the equation $\tau = \eta (\partial v / \partial y) $, where $\hat{y}$ is perpendicular to the plates.
Here, $\eta$ is the coefficient of shear viscosity, which is a measure of the efficiency of the momentum transfer in the medium.

In the nuclear context, many investigations concentrated on characteristics of giant resonances in order to infer the viscosity of nuclear matter (\cite{mondal_experimental_2017}, see references in \cite{auerbach_eta/s_2009}).
This relies on the validity of a hydrodynamic description down to zero temperature where the nucleon mean free path diverges.
We find that hydrodynamics fails to describe energetic reactions where the mean free path, while short, is not short enough for a hydrodynamic description to hold, requiring the use of transport theory to extrapolate to equilibrium or near-equilibrium situations.

Several groups have investigated the aforementioned $\eta / s$ ratio for different models utilized in nuclear collisions at intermediate energies, such as statistical multifragmentation \cite{pal_shear_2010} and quantum molecular dynamics (\textsc{QMD}) \cite{zhou_shear_2012}.
However, the latter investigations did not link viscosity to specific observables and did not aim at generality of the results beyond the specifics of the models.
Zhou \textit{et al.} noticed a correlation between shear viscosity and the strength of elliptic flow \cite{zhou_correlation_2014}. However, they did not validate that their model was accurately predicting viscosity-related observables by comparing to experimental data; therefore, their result is helpful for gaining a qualitative understanding of a theoretical relationship, but it is less reliable for learning about absolute bulk properties. Finally, the relaxation-time approaches \cite{xu_shear_2013-1, guo_isovector_2017} are suitable for order-of-magnitude estimates, but not for quantitative assessments.

The shear viscosity coefficient $\eta$, derived from the Boltzmann equation in Ref.~\cite{danielewicz_transport_1984} (see also \cite{shi_nuclear_2003}), is
\begin{equation}\label{eq:buu-viscosity}
\eta = \frac{5 T}{9} \frac{\left( \int \mathrm{d}\vec{p}_1 f_1 p_1^2 \right)^2} %
       {\int \mathrm{d}\vec{p}_1 \,\mathrm{d}\vec{p}_2 \,\mathrm{d}\Omega f_1^{} f_2^{} \tilde{f}_1' \tilde{f}_2'%
        v_{12} \frac{\mathrm{d}\sigma}{\mathrm{d}\Omega} q_{12}^4 \sin^2 \theta  } \, .
\end{equation}
Here, the elastic scattering cross section, $\mathrm{d}\sigma/\mathrm{d}\Omega$, is scaled with a factor $q^4 \sin^2 \theta$, which emphasizes large relative momenta $q$, where $q_{12} = \abs{\vec{p}_1 - \vec{p}_2}/2\,$, and wide scattering angles $\theta$. Thus, in kinetic transport, viscosity is tied to $\dfrac{d\sigma}{d\Omega} q^4 \sin^2 \theta$, sometimes called the ``transport cross section'' --- the differential particle-particle cross section scaled with a weight that increases with relative momentum and scattering angle. 
To learn about the shear viscosity, we will adjust the NN cross sections to match the stopping data, and we will draw conclusions about the viscosity using Eq.~\ref{eq:buu-viscosity}.

\section{The NN cross section in the nuclear medium}\label{sec:cs-reductions}

Looking ahead, comparisons to data clearly demonstrate that using the bare nucleon-nucleon cross section in the \textsc{BUU} equation (\ref{eqn:buu}) overestimates the amount of stopping found in central collisions at intermediate energy.
There are several different perspectives on the $\sigma_\mathrm{NN}$ in the medium.

Many groups follow the assumption that cross sections (CS) should scale with the nucleon effective mass \cite{pandharipande_nuclear_1992,persram_elliptic_2002,li_nucleon-nucleon_2005}.
This would require the nuclear transition matrix to stay the same in the medium as in vacuum, which is a perturbative approximation that does not hold for nuclear interactions. Therefore, there are questions about the validity of this assumption.
Further, the cross section should also be affected by the isospin asymmetry of the surrounding medium and several other factors not emphasized in the scaling.
In confronting the microscopic theory with the scaling \cite{sammarruca_microscopic_2014}, Sammaruca concluded that \underline{no} simple phenomenological ansatz following effective mass scaling is valid.
Some authors simply take \imnncs{} as a fraction, e.g.\ half, of the free cross section \cite{cai_yanhuang_semiclassical_1989,gaitanos_stopping_2004,zhou_thermodynamic_2013, basrak_aspects_2016}.
The deficiency of this assumption is that the free NN cross sections are not recovered when the matter becomes sufficiently dilute.
Following the transition matrix approach, one can derive, in the quasi-particle limit, both the mean field and in-medium cross section, making the development of the Boltzmann equation more self-consistent, in principle, changing both sides of Eq.~\ref{eqn:buu} \cite{alm_critical_1994, gaitanos_nuclear_2005}.
With this, though, the collision modification is only due to the mean field and statistics, and collisions do not affect each other.

In the early phenomenological parametrization of \imnncs{} in the literature, the cross section was assumed to change linearly with density \cite{westfall_mass_1993}. Eventually, with rise in density, this results in negative values. Another phenomenological approach was later adopted, where the cross section was assumed to reduce to a geometric unitary limit at high density \cite{danielewicz_hadronic_2002}.

In this paper, three scenarios for intermediate cross sections are discussed and then explored.
We resort to those scenarios because the free NN cross sections are found to be too large to describe data.
In the \textsc{BUU} simulations, the application of the \imnncs{} results in a reduced probability for NN collisions, compared to $\sigma_\textsc{NN}^\textrm{free}$.
For practical use in the simulations, each reduced cross section can be presented as a reduction factor multiplied by the free cross section.
That reduction factor gets sampled statistically in the simulations.
Pauli blocking of the final state is incorporated separately from this reduction factor.

\subsection{Tempered cross sections}

The Tempered cross section reduction scheme \cite{danielewicz_hadronic_2002} is arrived at by considering unitarity.
For a particle moving through a medium of number density $n$, the scattering partners are distributed at a relative distance of $\approx n^{-1/3}$. For two-body collisions to be independent from each other, the cross sections should be no larger than a value of the order of the distance squared, or $n^{-2/3}$,
\begin{equation}\label{eq:tempered-sigma0}
 \imnncs{} \lesssim \sigma_0 \equiv \nu n^{-2/3} \, ,
\end{equation}
where $\nu$ is of the order of 1. As the medium becomes more dilute, though, the cross sections should reach their free-space limit.
We parameterize the gradual change between the free and unitary limits with the formula
\begin{equation}\label{eqn:tempered}
 \imnncs{} = \sigma_0 \tanh \left( \frac{\sigma^\text{free}_\textsc{NN}}{\sigma_0} \right),
\end{equation}
where $\sigma_0$ is defined with the r.h.s.\ of Eq.~\ref{eq:tempered-sigma0} (in principle, other smooth interpolating functions could be used). As energies in an NN subsystem increase, the free NN cross sections become increasingly anisotropic, peaking in the forward and backward directions.
Those peaks are tied to higher angular momentum values. When particles are more tightly packed in a medium, these cross section contributions  should be more suppressed than contributions from lower angular momenta.
For the anisotropic cross sections, we adopt a modification of Eq.~\ref{eqn:tempered}:
\begin{equation}\label{eqn:temper-omega}
 \left( \frac{\mathrm{d}\sigma}{\mathrm{d}\Omega} \right)^\textrm{med}
 = \frac{\sigma_0}{4\pi}
 \tanh \left[ \frac{4\pi}{\sigma_0}
              \left( \frac{\mathrm{d}\sigma}{\mathrm{d}\Omega} \right)^\textrm{free}
      \right] \,.
\end{equation}
Here, for the purposes of the cross section, particles are treated as distinguishable, even when they belong to the same species.
The equation above accomplishes the goal of preferentially suppressing the forward and backward peaks, or contributions from high angular-momenta, relative to those from lower momenta.
With Eq.~\ref{eqn:temper-omega}, the cross sections become low and isotropic in the high-density limit, with the absolute limit on differential cross section of $\sigma_0 / 4\pi$. We stress here again that we treat the particles as distinguishable in the determination of cross section.

\subsection{Rostock cross sections}

Some early microscopic calculations of in-medium cross sections were carried out at the University of Rostock \cite{my_alm_critical_1994}, within a thermodynamic T-matrix approach.
In these calculations, the cross sections were modified to account for Pauli blocking in intermediate states and due to single-particle energy shifts \cite{alm_-medium_1995}.
Moreover, the results were derived assuming that the total momentum of the particles was zero in the frame of the local nuclear matter, in order to simplify the calculations.
We coarsely capture the essence of the results \cite{alm_critical_1994} with the following parametrization of the cross section reduction:
\begin{equation}
\imnncs{} = \sigma^\textrm{free}_\textsc{NN} \exp \left( -0.6 \frac{\rho/\rho_0}{1+\left[T_\text{c.m.} / (150\,\text{MeV})\right]^2} \right),
\end{equation}
where $T_\text{c.m.}$ is the total kinetic energy of the two interacting particles, in the frame where the local medium is at rest.

\subsection{Fuchs cross sections}

Fuchs \textit{et al.} \cite{fuchs_off-shell_2001} underscored that in the \textsc{BUU} equation (\ref{eqn:buu}), the in-medium mean fields that the particles are subject to on the l.h.s.\ of the equation should be derived consistently with the in-medium NN cross sections $\sigma$ used in the collision integral on the r.h.s.
As the basis for these simultaneous alterations, they employed the in-medium Dirac-Brueckner T-matrix \cite{fuchs_off-shell_2001}.
Like the Rostock one, this cross section was derived for two particles with total momentum equal to zero in the rest frame of the local medium.
The cross section reduction of Fuchs \textit{et al.} is parameterized here with

\begin{equation}
\sigma_\textrm{nn}^\textrm{med} = \sigma_\text{nn}^\textrm{free}
 \exp \left( -1.7 \frac{\rho/\rho_0}{1+\left[T_\text{c.m.}/ (12\,\text{MeV})\right]^{3/2}} \right)
\end{equation}
\begin{equation}
\sigma_\text{np}^\textrm{med} = \sigma_\text{np}^\textrm{free}
                   \exp \left( -1.4 \frac{\rho/\rho_0}{1+[T_\text{c.m.} / (33\,\text{MeV})]} \right).
\end{equation}

The different cross section reductions above will be used to compare predictions from pBUU to stopping data, to see if stopping is sensitive to the absolute or to the transport cross section similarly to the shear viscosity. If the latter dependence were monotonic, then we could tune the NN cross section to reproduce experimentally observed stopping and use that tuned cross section to calculate the shear viscosity self-consistently.

\section{Stopping observables}

Different observables have been used in the literature to characterize nuclear stopping. The observables tend to be optimized for a specific energy range where measurements are carried out. We use several of these observables to enable larger energy range coverage and better discern the robustness of our conclusions.

\subsection{Linear momentum transfer}

\begin{figure}
    \begin{center}
    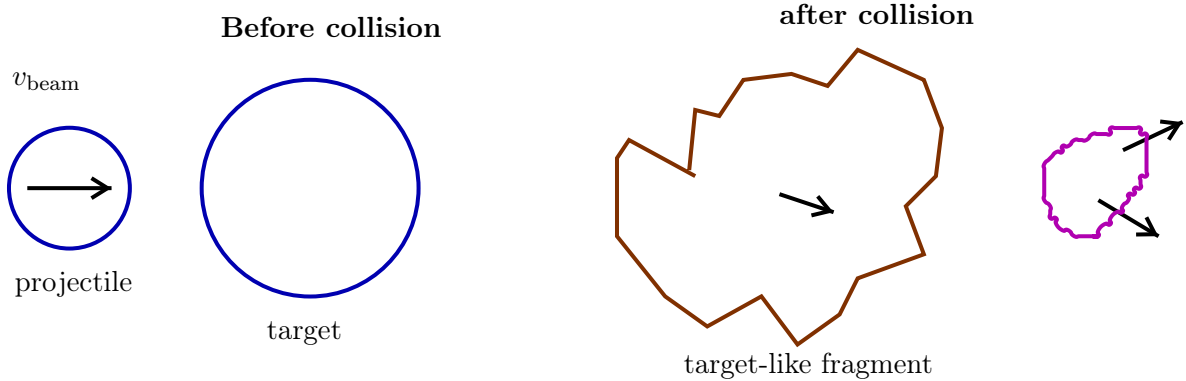
    \end{center}
    \caption{
    Schematic of asymmetric collision.
    Projectile transfers momentum to the target.
    To assess linear momentum transfer, the longitudinal velocity of the target-like fragment is compared to the velocity of the center-of-mass of the collision.}\label{fig:lmt-schematic}
\end{figure}

In a mass-asymmetric collision of a light projectile colliding with a heavy target, one can assess the momentum that is transferred to the target, and thus have a measure of the stopping power --- that is, a reflection of how much the projectile decelerates when it interacts with the target, provided the target survives in some form.
As in the schematic in Fig.~\ref{fig:lmt-schematic}, one finds the laboratory-frame velocity of the largest fragment emitted from the collision, assuming that this fragment stands out.
Because of the high mass, that fragment is assumed to originate from the target (the ``target-like fragment'').
The higher its velocity, the more momentum was transferred from the projectile.
This corresponds to a higher degree of stopping.
To compare the observable across reaction systems, this fragment velocity is divided by that of the center of mass.
Since the velocities involved are non-relativistic, they can be used to infer the scaled linear momentum transfer (\textsc{LMT}). This observable was originally used to distinguish between direct and compound fission reactions in heavy nuclei \cite{nicholson_direct-interaction_1959,sikkeland_momentum_1962}, then used more generally in nucleus-nucleus collisions \cite{viola_linear_1982}.

The technique relies on a clear determination of the target-like fragment, and as the beam energy increases, there are more violent collisions and the largest fragment produced becomes lighter.
In consequence, the practical energy range for this observable is from around the Coulomb barrier to around 150$\,$MeV/nucleon or so.
Above this range, any fragment that could be tied to the target is difficult to distinguish from other fragments of similar intermediate mass.

The observable \textsc{LMT} is defined \cite{colin_splintering_1998} as
\begin{equation}
\textsc{LMT} = \left\langle \frac{v_\parallel}{v_\text{c.m.}} \right\rangle \, ,
\end{equation}
where $v_\parallel$ is the velocity of the target-like residue in the beam direction, $v_\mathrm{c.m.}$ is the velocity of the reaction center of mass. A higher \textsc{LMT} corresponds to a higher degree of stopping.

\begin{figure}
    \begin{center}
        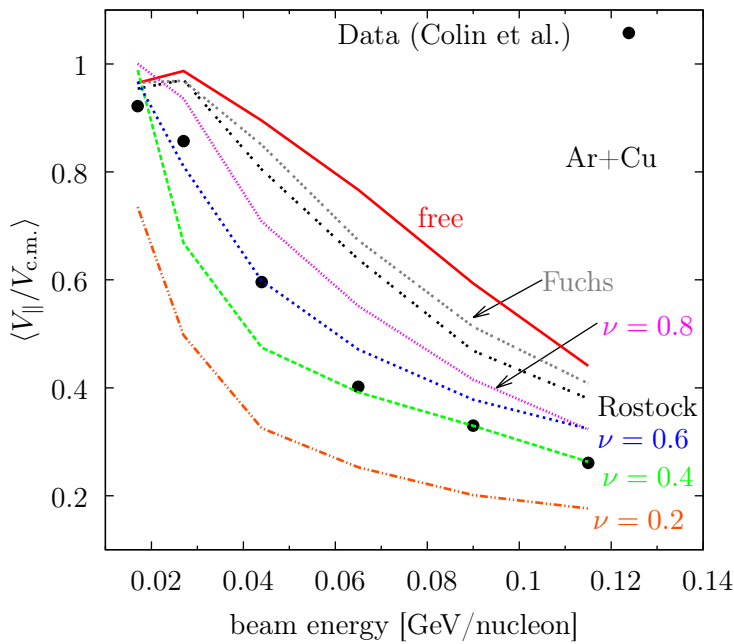
        \caption{Linear momentum transfer for $^{40}$Ar+$^{}$Cu.
        Lines represent the theoretical results incorporating different in-medium NN cross sections.
        The ``Tempered'' reductions are marked with their adjustable parameter $\nu$.
        Symbols represent experimental data \protect\cite{my_colin_splintering_1998}.}\label{fig:lmt-arcu}
    \end{center}
\end{figure}

\begin{figure}
 \begin{center}
    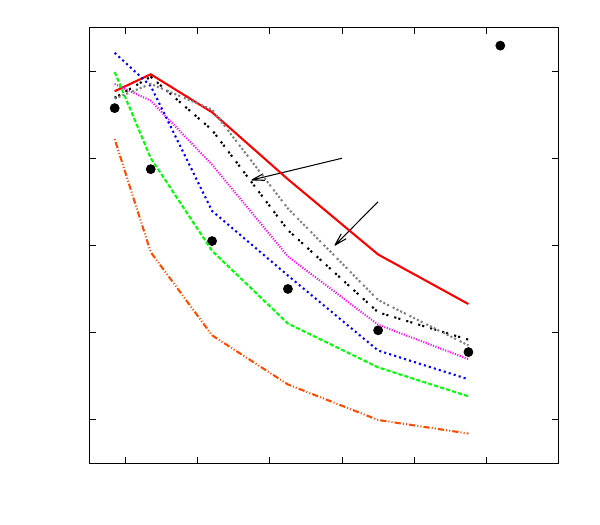
    \caption{Linear momentum transfer for $^{40}$Ar+Ag. Lines represent the theoretical results incorporating different in-medium NN cross sections. The ``Tempered''
        reductions are marked with their adjustable parameter
        $\nu$. Symbols represent experimental data \protect\cite{my_colin_splintering_1998}.}\label{fig:lmt-arag}
 \end{center}
\end{figure}

\begin{figure}
    \begin{center}
        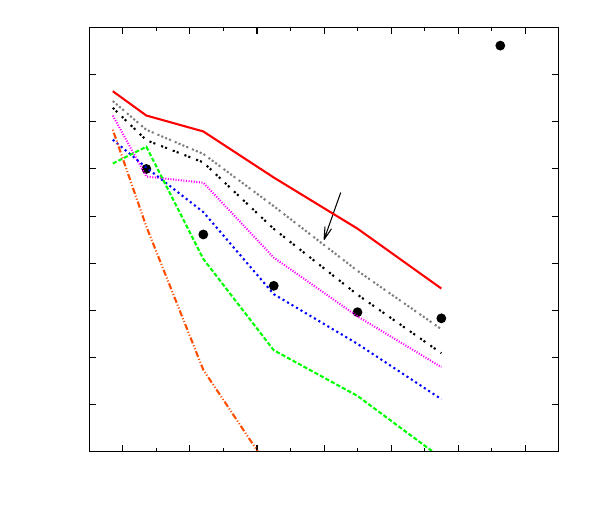
        \caption{Linear momentum transfer for $^{40}$Ar+Au. Lines represent the theoretical results incorporating different in-medium NN cross sections. The ``Tempered''
            reductions are marked with their adjustable parameter
            $\nu$. Symbols represent experimental data \protect\cite{my_colin_splintering_1998}.}\label{fig:lmt-arau}
    \end{center}
\end{figure}

Experimental \cite{colin_splintering_1998} and theoretical results for \textsc{LMT} in collisions of a $^{40}$Ar projectile with Cu, Ag, and Au targets are shown in Figs.~\ref{fig:lmt-arcu}, \ref{fig:lmt-arag}, and \ref{fig:lmt-arau}, respectively.
At low energies, \textsc{LMT}$\simeq 1$, indicating formation of a compound system and complete stopping. As beam energy increases, transparency sets in and \textsc{LMT} decreases. 
In the experiment, it appears that targets were used with their natural isotopic content.
To determine $V_\parallel$ in the equation above, a filter on just the heaviest fragments was used, with the assumption that these heaviest fragments provided a good average estimate of the longitudinal momentum of the target remnant.

In the \textsc{BUU} calculations, we use the specific isotopes $^{63}$Cu, $^{107}$Ag, and $^{197}$Au for the targets. In central collisions at this energy, the one- or two-neutron differences in the target content should not impact \textsc{LMT} enough to matter. Within our simulation, the target remnant is explicitly tracked throughout the collision, and its velocity is directly calculated from the constituent particles.
In particular, nucleons that initially belonged to the target and continue to be bound are considered to be part of the target remnant.
To be considered bound, the particle energy must be at least $6\,$MeV below the continuum in the local frame.  For charged particles, the energy excludes the Coulomb contribution, i.e.\ the continuum is counted from the top of the local Coulomb barrier.

The various \imnncs{} schemes described in Section~\ref{sec:cs-reductions} are tested in \textsc{BUU} calculations, with the corresponding results shown with lines in Figs.~\ref{fig:lmt-arcu}--\ref{fig:lmt-arau} alongside the experimental data.
The Rostock and Fuchs reductions, as well as the case with no reduction (``free''), are labeled with their names, while the Tempered CS is marked by its tunable parameter $\nu$.
It is clear from the \textsc{LMT} figures that the free cross section overestimates the stopping in all three reaction systems, and that the Tempered CS with $\nu=0.2$ underestimates it.
The Rostock and Fuchs reductions produce \textsc{LMT} values that are very close to each other in all cases, with Fuchs resulting in $\sim7\%$ higher values than Rostock in the 65$\,$MeV/nucleon region.

Generally, use of the free cross section results in a coarsely linear dependence of \textsc{LMT} on beam energy in all three systems at about 27$\,$MeV and higher, while the reductions all exhibit a positive concavity with energy in the $^{40}$Ar$\,+\,$Cu and $^{40}$Ar$\,+\,$Ag cases, which more closely resembles the data.
In the case of $^{40}$Ar$\,+\,$Au, all calculated lines show a roughly linear dependence on beam energy, while the experimental data shows an even larger concavity compared to the lighter systems.

Judging by eye, the cross section that best fits the $^{40}$Ar$\,+\,$Cu and $^{40}$Ar$\,+\,$Ag data is the Tempered one with $\nu=0.4$ or 0.6.
In the $^{40}$Ar$\,+\,$Au reaction, the cross section that best fits the data seems to be the Tempered CS with $\nu=0.8$.

\subsection{Rapidity variance ratio}

If particles in the hot, dense region of a nuclear collision undergo many collisions (because the mean free path becomes comparable to the typical interparticle distance), the region tends to equilibrate, and particles will lose memory of which direction they were originally traveling in.
With this, more stopping will occur and emission from this specific region will tend towards isotropy in the reaction center of mass.
The FOPI Collaboration provides a practical measure of this isotropy with the observable \textit{varxz}, defined as \cite{reisdorf_systematics_2007}
\begin{equation}
 \textit{varxz} = \frac{\Delta y_x}{\Delta y_z},
\end{equation}
where $\Delta y_x$ is the variance of particle rapidity along a randomly chosen direction that is transverse to the beam and $\Delta y_z$ is the variance of the standard particle longitudinal rapidity.
 
\begin{figure}
    \begin{center}
        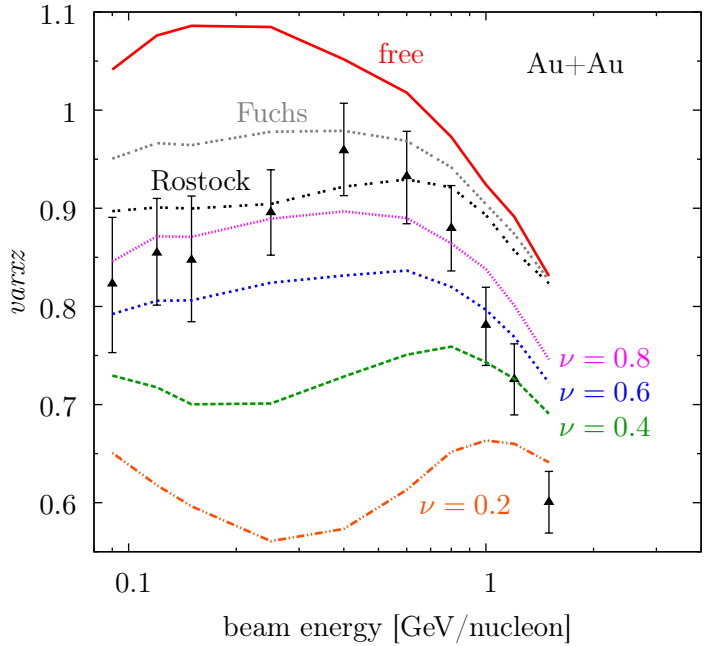
        \caption{Stopping observable \textit{varxz} for protons in \coll{Au}{Au} collisions at different beam energies at $b_\text{red}<0.15$.
Lines show the effects of different in-medium reductions of the NN cross section.
The ``Tempered'' cross-section reductions are marked with their tunable parameter $\nu$.
Symbols are experimental data from the \textsc{FOPI} Collaboration. \protect\cite{my_reisdorf_systematics_2010}.}\label{fig:varxz-nocomp-au}
    \end{center}
\end{figure}

Fig.~\ref{fig:varxz-nocomp-au} shows the experimental results from \textsc{FOPI} \cite{reisdorf_systematics_2010} as well as the p\textsc{BUU} transport simulation results with the various \imnncs{} reduction schemes used, for \coll{Au}{Au}, looking at the distribution of protons. The experiment was carried out using the heavy ion accelerator \textsc{SIS} at \textsc{GSI}/Darmstadt, and charged particles were detected with a good coverage of angles throughout the $4\pi$ region, using the \textsc{FOPI} detector and a set of other detectors that provided particle tracking, energy loss determinations, time of flight determination, and charged particle identification. The beam energies spanned the range from 0.09 to 1.5$\,$GeV/nucleon, and the impact parameter selection was limited to $b_{\textrm{red}} \equiv b/b_{\textrm{max}} \lesssim 0.15$.

The most startling finding in Fig.~\ref{fig:varxz-nocomp-au} is that simulations with free cross sections yield \textit{varxz} clearly in excess of 1, in a wide energy range of 0.09 -- 0.6 GeV/nucleon, while \textit{varxz} seems to always stay below 1 in measurements.
This clearly eliminates the possibility of cross sections staying the same in the medium as in free space.
Thus, just like LMT, the \textit{varxz} comparison points to a reduction of the cross sections in the medium.
In the simulations with free cross sections, the matter exhibits a strong hydrodynamic behavior, splashing to the sides in central collisions \cite{danielewicz_effects_1995}, yielding $\textit{varxz}>1$.
On the other hand, in the measurements, even the isotropy is never reached, with the medium always staying partially transparent.

Both in the calculations with free cross sections and in the measurements, \textit{varxz} eventually drops as energy increases.
This can be attributed to two factors: typical momenta in the center of mass become large compared to the Fermi momentum, and cross sections become increasingly isotropic with the increase in energy.
Additionally, as energy increases, inelastic processes give rise to $\Delta$ resonances and pions.
The higher the energy, the more important those inelastic processes become.
While adjusting the in-medium cross sections, we can arrive at similar \textit{varxz} values as in experiment at energies below 0.8 GeV/nucleon.
At higher bombarding energies, the theoretical results with different cross sections begin to merge and exceed experimental \textit{varxz}.
This is because, in the simulations, we adjust only elastic cross sections and leave inelastic intact, and the balance in the importance shifts to the latter cross sections as energy increases.

When examining the p\textsc{BUU} simulations with the tempered \imnncs{} at $\nu=0.6$ (a reasonable parameter), at energies from 90 to $1500\,$MeV/nucleon, the ratio of peak $\Delta$ production and absorption rates, which are the primary inelastic processes, to peak elastic collision rates varies from 0 to 0.8. Assuming that the inelastic collisions start significantly affecting the reaction dynamics when the ratio is about 0.2 or 0.3, then we should look at beam energies of less than $~600\,$MeV/nucleon in deciding on in-medium cross sections.
Given this caveat, it seems the \imnncs{} that best fits the data below 600$\,$MeV/nucleon is either $\nu=0.8$ or Rostock.

Another caveat concerning conclusions on in-medium cross sections concerns another type of inelastic process, namely cluster production and breakup.
The clusters get more copiously produced in colder regions of the matter and the production predominantly takes place at subnormal densities.
In the pBUU simulations, we have the option of activating the production of $A = 2$ and $A = 3$ fragments.
The production is limited in the simulations to the densities $\rho \lesssim 0.6 \rho_0$, but the production and breakup rates are calculated based on processes taking place in the vacuum without in-medium modifications. As the processes of cluster production and breakup can compete with two-body collisions, there may be concerns about the ability to make conclusions about the medium modifications of the two-body processes.

Most often, due to concerns about the impact of any double-counting of interactions, we carry out the calculations of stopping with the cluster production switched off. However, for the sake of testing the validity of concerns tied to the cluster production and absorption, we also carried out calculations of Au+Au reactions with the cluster production activated.
The peak region in \textit{varxz} vs.\ energy, where the comparison between data and theory may be most telling, seems to be generally best described with the Rostock cross section or the tempered cross section with $\nu = 0.8$, regardless of whether the calculation includes cluster production or not.

\begin{figure}
 \begin{center}
  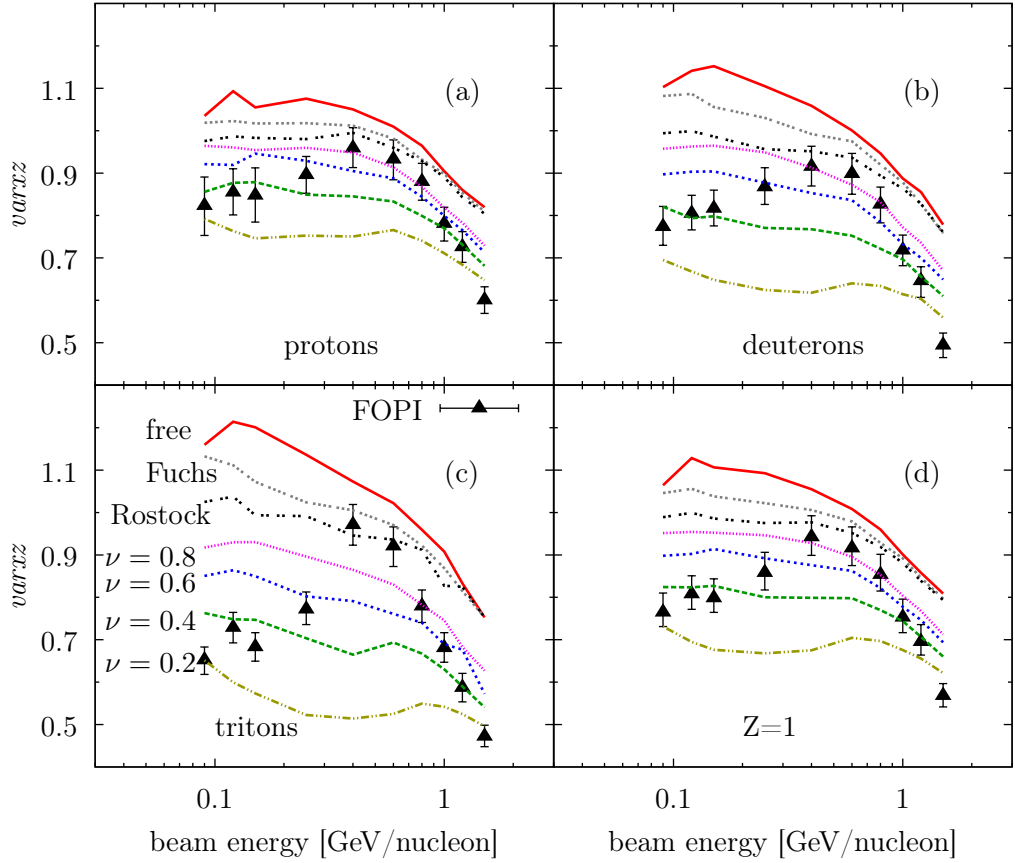
  \caption{Analogous to Fig.~\ref{fig:varxz-nocomp-au}, except that $A = 2, 3$ composite particle formation is enabled in the simulations, so a comparison to results for different particle species in the experiment \cite{reisdorf_systematics_2010} becomes possible.}\label{fig:varxz-comp-au-species}
 \end{center}
\end{figure}

\subsection{Isospin tracer}

Another observable that we use for assessing cross sections, the so-called isospin tracer, attempts to identify the relative yield of particles from the target and projectile in a given region of momentum space by examining isospin content.
This is done by studying collisions between nuclei with identical mass number but different charge number, interchanging the projectile and target roles, and then comparing the results to those from collisions of identical nuclei.

The method is described in more detail here:
\begin{quotation}
    ``The ($N/Z$)-tracer method is based on the following idea:
    let us assume that we are observing the final number of
    protons, $Z$ in a given cell of the momentum space. The
    expected yield $Z^\text{Ru}$ measured for the Ru $+$ Ru reaction
    is higher than $Z^\text{Zr}$ of the \coll{Zr}{Zr} reaction since Ru has
    44 protons as opposed to 40 for Zr. Such measurements
    using identical projectile and target deliver calibration values
     $Z^\text{Ru}$ and $Z^\text{Zr}$ for each observed cell. In the case of a
    mixed reaction, \coll{Ru}{Zr} or \coll{Zr}{Ru}, the measured proton
    yield $Z$ takes values intermediate between the calibration
    values ($Z^\text{Ru}$, $Z^\text{Zr}$). If, e.g., $Z$ is close to $Z^\text{Ru}$ in a \coll{Ru}{Zr}
    reaction, means that the cell is populated predominantly
    from nucleons of the Ru projectile while if it is close to $Z^\text{Zr}$
    it is mostly populated from nucleons of the Zr target. In
    this way it is possible to trace back the relative abundance
    of target to projectile nucleons contributing to a given cell.'' \cite{rami_isospin_2000}
\end{quotation}
Within the method, one constructs the observable $R_Z$, defined as
\begin{equation}\label{eq:rz}
R_Z = \frac{2 \times Z - Z^\text{Zr} - Z^\text{Ru}}{Z^\text{Zr} - Z^\text{Ru}} \, ,
\end{equation}
which assesses relative abundances of the projectile-target nucleons. In this case, $Z$ represents proton yield in a reaction with different projectile and target for a given location in momentum space. Yields for other particles can also be used in velocity space \cite{rami_isospin_2000}.
With the above definition, one arrives at $R_Z = 1\: (-1)$ when a momentum cell gets populated by protons originating exclusively from
 the Zr (Ru) nucleus, as long as the dynamics do not depend on the charge content.
The case of complete stopping would mean that the protons completely mix and, for any cell, half come from Zr and half from Ru. Thus, full
stopping is expected to yield $R_Z \equiv 0$.

\begin{figure}
    \begin{center}
        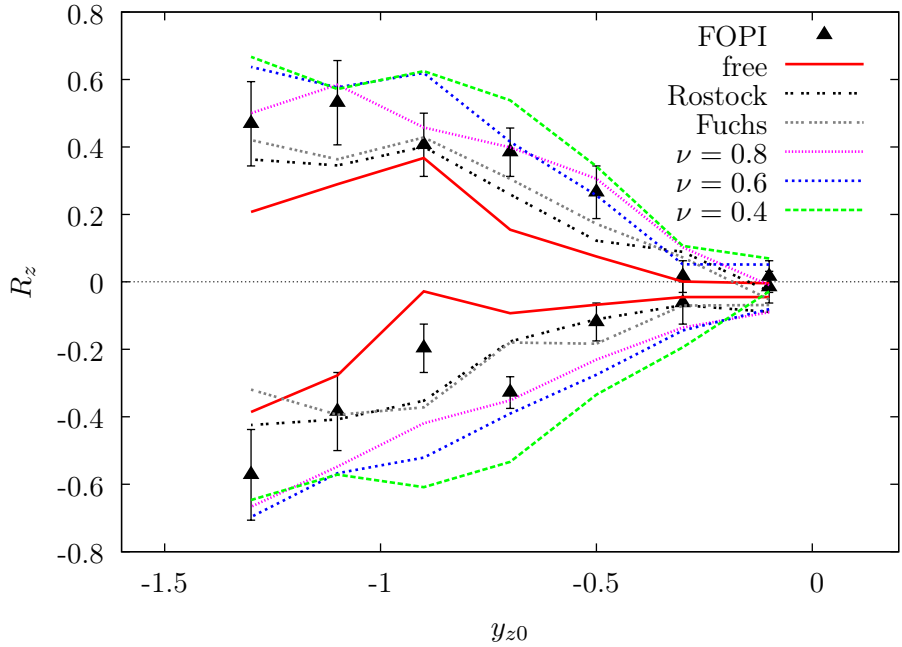
        \caption{Isospin tracer observable for central collisions of \collpt{96}{Zr}{96}{Ru} (bottom) and \collpt{96}{Ru}{96}{Zr} (top) at 400$\,$MeV/nucleon vs.\ scaled center-of-mass rapidity, such that $y_{z0}=-1$ is the initial projectile rapidity.
Data is from the FOPI collaboration \protect\cite{rami_isospin_2000}. A tendency of $R_z$ to stay closer to zero at finite rapidities indicates a higher degree of mixing, and thus stopping.}\label{fig:rz-nc}
    \end{center}
\end{figure}

The experimental results \cite{rami_isospin_2000}, along with results from the \textsc{BUU} transport model, with the $Z$ in $R_Z$ representing proton yield, for collisions between $^{96}$Zr and one of its $A=96$ counterparts, $^{96}$Ru, are shown in Fig.~\ref{fig:rz-nc} plotted against rapidity, for beam energy of 400$\,$MeV/nucleon and $b_\text{red}<0.12$.

As rapidity (in the center of mass) gets more negative, the momentum cells are increasingly populated by protons from the target, according to $R_Z$ and the interpretation above.
This makes sense, as the target's particles are more likely to persist in the backward rapidity region for limited momentum transfers in interactions.
As the bins closer to midrapidity are examined, it is seen that those bins get populated by protons from both colliding nuclei, as $R_Z$ is close to zero there.

It is again clear from Fig.~\ref{fig:rz-nc} that use of the free \textsc{NN} cross section overestimates the stopping or mixing in this case, when assessed with the $R_Z$ observable.
It seems that the \imnncs{} best fitting the data here is either Rostock, Fuchs, or Tempered with $\nu \sim 0.8$.
The $R_Z$ observable is challenging for Monte Carlo calculations, due to statistical fluctuations that get emphasized in the subtraction of similar values, $Z^\text{Zr} - Z^\text{Ru}$, and further amplified in the division by the resulting small number in Eq.~\ref{eq:rz}.

\subsection{Summary of in-medium cross section analysis}

A summary of the stopping observables that were investigated in different systems and the optimal \imnncs{} for each is given in Table~\ref{tab:stopping}. Overall, there is not one \imnncs{} that optimally reproduces the stopping across all observables, sizes, and energies.
It is clear, though, that the cross section is reduced. For the remaining investigations, we use Tempered with $\nu=0.6$ as it is the most representative of the range of conclusions.

\begin{table}
 \begin{tabular}{l|c|c|c}
  observable & reaction system & energies [MeV] & best cross section reduction \\
  \hline
  \textsc{LMT} & $^{40}$Ar$\,+\,$Cu  & 17--115 & Tempered w/ $0.4 \lesssim \nu \lesssim 0.6$ \\
  \textsc{LMT} & $^{40}$Ar$\,+\,$Ag  & 17--115 & Tempered w/ $0.4 \lesssim \nu \lesssim 0.6$ \\
  \textsc{LMT} & $^{40}$Ar$\,+\,$Au  & 27--115 & Tempered w/ $\nu \simeq 0.8$ \\
  \textit{varxz} & Au$\,+\,$Au & 90--1500 & Tempered w/ $\nu \simeq 0.8$ or Rostock \\
  $R_z$ & $^{96}$Zr+$^{96}$Ru & 400 & Tempered w/ $\nu \simeq 0.8$, Rostock, or Fuchs \\
  & (and inverse) & &
 \end{tabular}
 \caption{Summary of cross section determination results. No single cross section reduction is favored universally. The Tempered cross section with $\nu \sim 0.6$ is deemed to be the best compromise.}\label{tab:stopping}
\end{table}

As the last issue of potential concern in drawing conclusions from stopping observables, we discuss possible competition between cross sections and mean fields in shaping those observables.

\subsection{Mean-Field Sensitivity}\label{sec:sensitivity-eos}

Choices made regarding the energy functional can make the matter less or more compressible. For more incompressible matter, the mean field potentials become more quickly repulsive with increase in net density $\rho$. The mean field potentials can also depend on momentum $p$ in the local rest frame. The incompressibility of matter is commonly described in terms of the constant \cite{blaizot_microscopic_1995}
\begin{equation}
 K = 9 \rho_0 \left. \frac{\partial^2 (E/A)}{{\partial \rho}^2} \right|_{\rho=\rho_0} \, ,
\end{equation}
where $E/A$ is the energy per nucleon in cold symmetric matter, and the derivative is evaluated at the normal density $\rho_0$.
The calculations so far were all done employing a relatively conventional functional yielding incompressibility $K = 210\:$MeV and effective mass at Fermi momentum at $\rho_0$ of $m^*/m = 0.7$, where the latter characterizes the momentum dependence of $U$.
Some uncertainty regarding the incompressibility and momentum dependence remains, though, and the FOPI Collaboration \cite{reisdorf_systematics_2010} found some sensitivity of the stopping to the decisions made on the mean-field interactions in the Isospin Quantum Molecular Dynamics (IQMD) model \cite{hartnack_modelling_1998}.
Here we test whether we can observe any similar sensitivity.
An excessive sensitivity would hamper the efforts to learn about the in-medium cross sections.

\begin{figure}
\begin{center}
 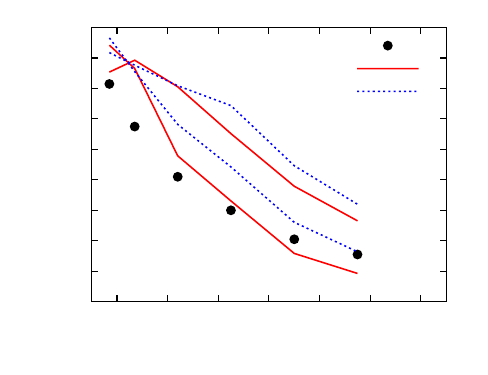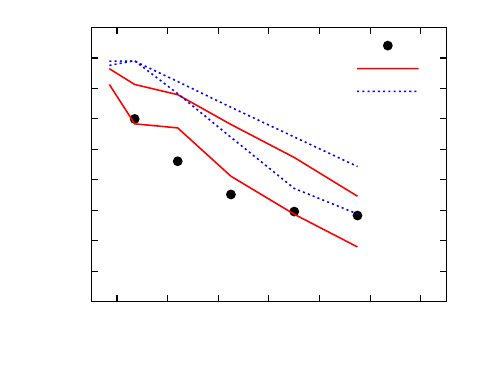
 (a) linear momentum transfer

\hspace*{\fill}
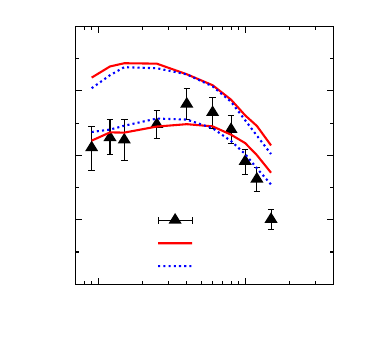\hspace{\fill}
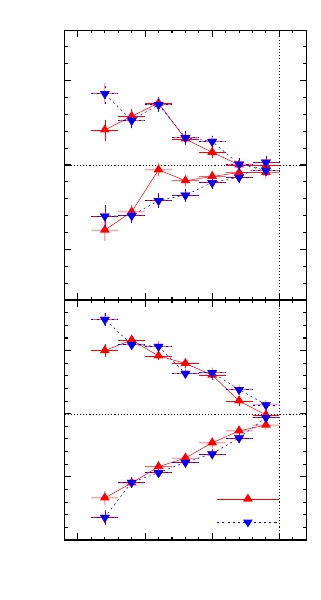\hspace*{\fill}
  
  \hspace*{\fill} (b) \textit{varxz} \hspace{\fill} (c) isospin tracer \hspace*{\fill}
 \caption{Mean-field sensitivity of stopping observables. S (H) refers to a soft (hard) compressibility, while M refers to the
 inclusion of momentum-dependence in the mean field. \textsc{SM} and \textsc{H} are limiting combinations that give similar predictions for flow in semicentral collisions \cite{pan_sideward_1993}.
 }\label{fig:eos-effects}
 \end{center}
\end{figure}

To test sensitivity to the mean field, we show in Fig.~\ref{fig:eos-effects} results obtained for stopping observables when using our standard mean field (soft, momentum-dependent, or ``SM''), corresponding to incompressibility $K=210\:$MeV, as well as a mean field with no momentum dependence, corresponding to incompressibility $K=380\:$MeV (hard, or ``H''). The two mean fields yield similar results for flow in semicentral collisions \cite{danielewicz_determination_2002}. However, the momentum-dependent mean field fails to explain flow at high impact parameters or high transverse momenta \cite{danielewicz_determination_2000}. While the H mean field is not realistic, its use allows us to assess the general sensitivity of the stopping observable to the choice of mean field.

Surprisingly, even when a quite extreme mean field like H is used, the stopping observables at high energies, \textit{varxz} and isospin tracer $R_Z$, change very little, suggesting relative robustness of conclusions there. We find some sensitivity to the mean field at lower energies, in the stopping observable LMT. Interestingly, no matter what mean field is used, the need for in-medium cross section modifications is apparent, in order to match the data in Fig.~\ref{fig:eos-effects}. The reduction in the momentum dependence in $U$, accompanied by an increase in the incompressibility to meet flow data from semicentral collisions, results in an enhanced stopping when judging that stopping with LMT. With this, to meet the data with reduced momentum dependence in $U$, one would need a stronger reduction in in-medium cross section. However, realistically, the uncertainty in $U$ and in incompressibility spans approximately a third of the range between SM and H. Given Figs.~\ref{fig:lmt-arau} and \ref{fig:eos-effects}, any deemed change in the reduction for in-medium cross section would be small compared to the ambivalence we already have.

\section{Stopping and the transport cross section}

\begin{figure}
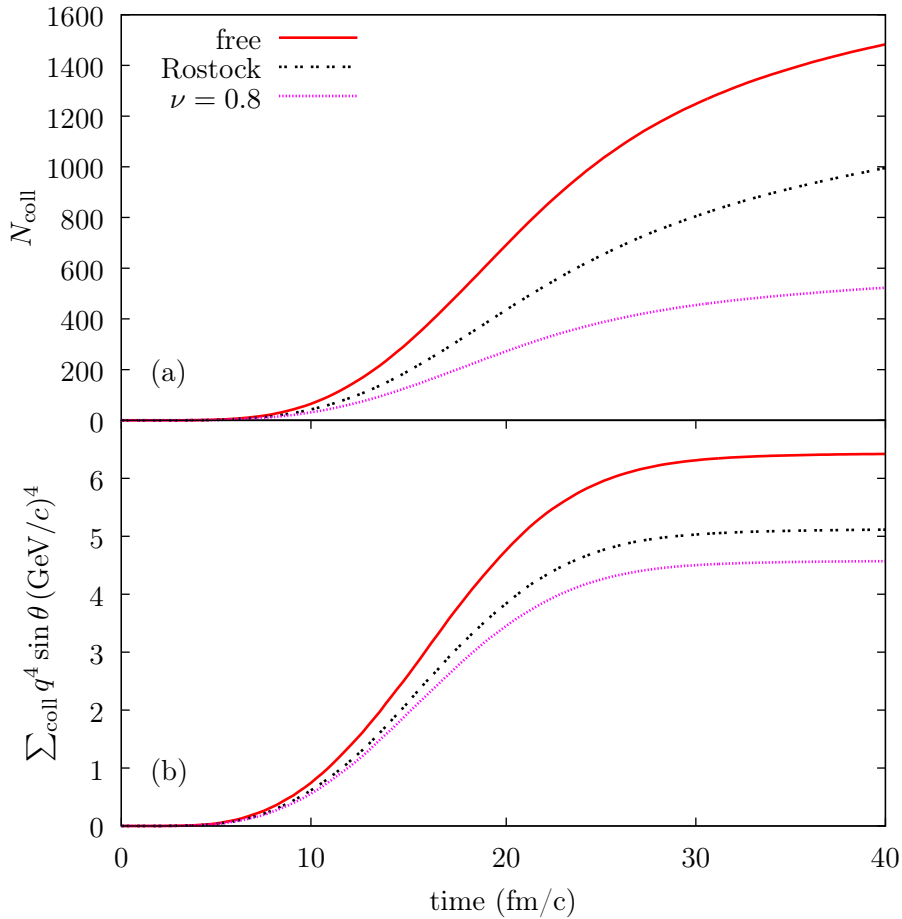
\caption{Cumulative number of elastic NN collisions vs.\ elapsed time in central \coll{Au}{Au} collisions at $400\,$MeV/nucleon, for three different in-medium cross sections. Panels (a) and (b) show, respectively, net number of collisions and number of collisions weighted with the viscous weight.}\label{fig:collsvstime}
\end{figure}

In-medium cross sections are obviously  not directly observable, and they are tied to the transport equation that relies on the concept of quasiparticles, which brings in a level of phenomenology.
So a question might be asked whether more robust conclusions may be drawn from the studies of stopping that extend beyond the cross sections.
To illustrate the precarious nature of the conclusions on cross sections, we show in Fig.~\ref{fig:collsvstime}a the collision number in \coll{Au}{Au} collisions at $400\,$MeV/nucleon obtained in simulations, with three different in-medium cross sections: free-space, Rostock, and tempered with $\nu=0.8$.
The stopping is similar in those collisions for the Rostock and $\nu=0.8$ cross sections, when quantified in terms of \textit{varxz}, and significantly reduced compared to free cross sections, as seen in Fig.~\ref{fig:varxz-nocomp-au}. Yet, in spite of the similar stopping in that figure, the collision count for the two cross sections is different by a factor of 2 in Fig.~\ref{fig:collsvstime}a. Apparently, the stopping does not directly correlate to the typical elementary cross section in a reaction, which in turn does not bode well for reaching physics conclusions from stopping.

Taking another perspective, the collision count includes some collisions that are hard, occurring at high relative velocity with large momentum transfer, and some that are soft, occurring at low relative velocity and low momentum transfer.
Those soft collisions contribute little to momentum transfer across the system as represented by observables such as LMT, \textit{varxz}, or $R_Z$.
Instead, one can consider that the most elementary macroscopic characteristics for a system, which are tied to cross sections, are transport coefficients.
The one tied directly to momentum transfer is the shear viscosity, and it involves the so-called transport cross sections.

\begin{figure}
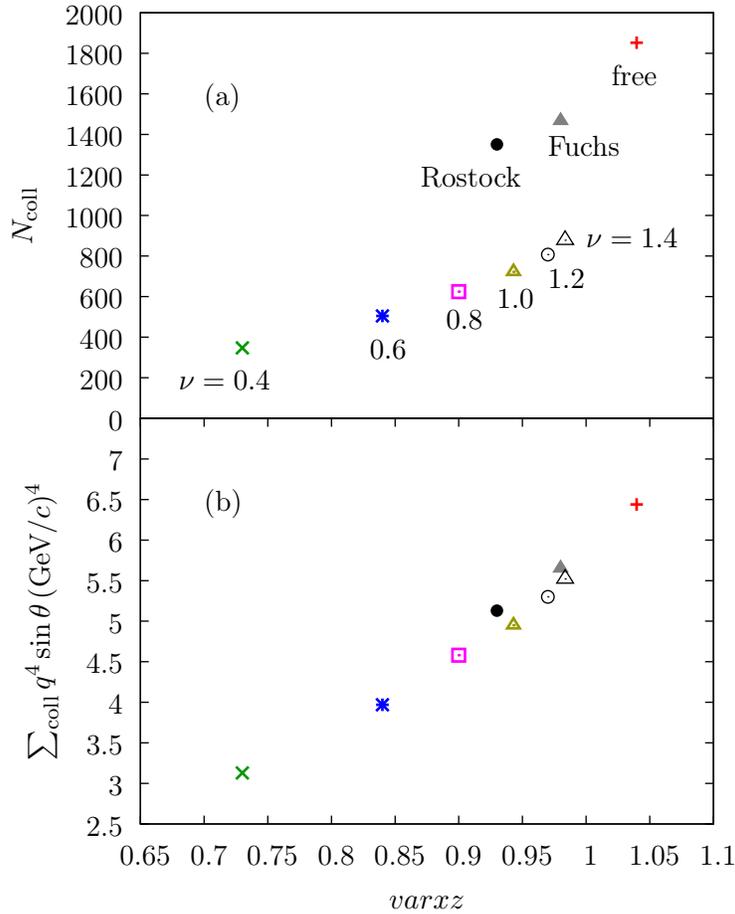
\caption{Correlation between the number of nucleon-nucleon collisions that took place up until $100\,$fm$/c$ and the stopping observable $varxz$ for the central $400\,$MeV/nucleon collision. The top panel utilizes the unweighted collision number, and the bottom panel utilizes the number weighted with the viscous weight. The weighted collision number rises monotonically with the stopping, unlike the unweighted number.}\label{fig:collvstop}
\end{figure}

The transport-type cross section may be recognized in Eq.~\ref{eq:buu-viscosity}. Here, the shear viscosity coefficient $\eta$ is dependent on the NN cross section, with the cross section's weight dependent on the relative momentum and scattering angle. In Fig.~\ref{fig:collsvstime}b, we show the weighted collision count, with each collision multiplied by its viscous weight factor $q^4 \sin^2 \theta$. The weighted collision count is similar for the Rostock and $\nu=0.8$ cross sections, consistent with \textit{varxz} values being similar for those two cross sections in Fig.~\ref{fig:varxz-nocomp-au}.
To provide more insight, in Fig.~\ref{fig:collsvstime} we plot the unweighted (top) and weighted (bottom) collision counts up to $100\,$fm$/c$ vs.\ \textit{varxz}, for a variety of \imnncs{}. While the stopping correlates with the unweighted collision count, the correlation is fairly broad, with the count differing by up to a factor of 2 for some different plausible cross section reductions.
Here, one can see that the stopping poorly tests the overall number of particle-particle collisions. However, the correlation of the stopping is fairly tight with the collision count when the collisions are weighted with the viscous weight, as seen in Fig.~\ref{fig:collvstop}. The broader systematics further support the view that the stopping tests transport cross sections and more broadly the medium shear viscosity.

\subsection{Viscosity from \textsc{BUU}}\label{sec:viscosity}

As Eq.~\ref{eq:buu-viscosity} was derived using the same assumptions as the transport model used to constrain the \imnncs{}, that cross section can be inserted into this equation to find a viscosity coefficient $\eta$ that is hopefully of greater generality than even the transport model itself, given the correlation shown in, for example, Fig.~\ref{fig:collvstop}. The calculation is performed with the effective mass described in Section~\ref{sec:sensitivity-eos}, which tends to increase the viscosity somewhat, compared to using the free mass.

\begin{figure}
 \begin{center}
  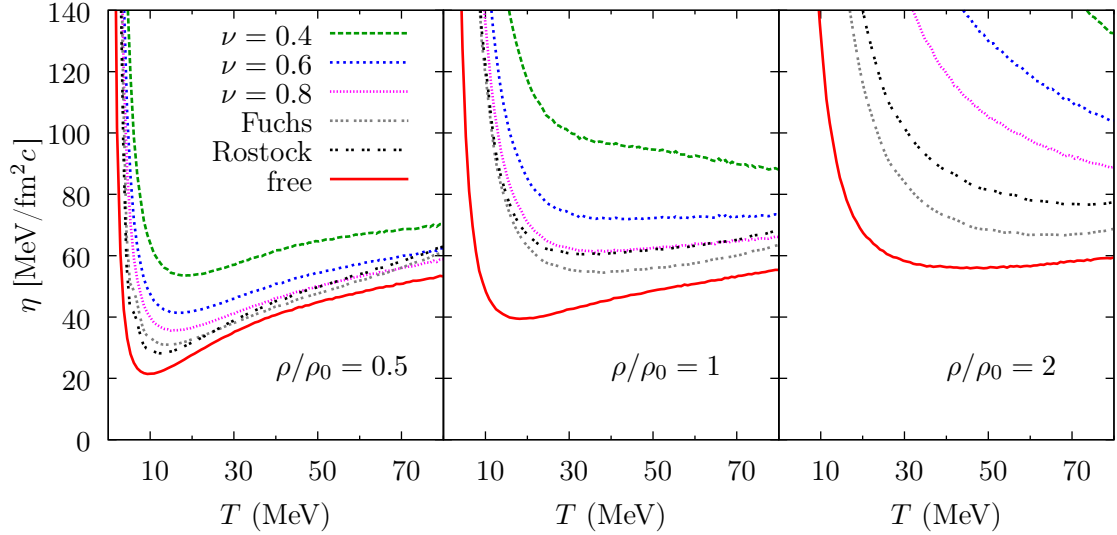
 \end{center}
 \caption{Shear viscosity for symmetric nuclear matter at different temperatures and densities, deduced from Boltzmann equation with in-medium cross sections.}\label{fig:viscos}
\end{figure}

The results of viscosity calculations are displayed in Fig.~\ref{fig:viscos}.
At all densities and cross section reductions presented, the viscosity grows indiscriminately at low temperatures.
This occurs when the nucleon system becomes degenerate and collisions become strongly Pauli-suppressed, with the weighted rate in the denominator in Eq.~\ref{eq:buu-viscosity} tending towards zero.
As temperatures increase and Pauli effects diminish, the collisions become more frequent.
The viscosity goes through a minimum and at high temperatures, it behaves in a classical fashion, growing like $\sqrt{T}$.
Eventually, inelastic processes set in and calculation of viscosity using just elastic processes in Eq.~\ref{eq:buu-viscosity} will start overestimating the actual viscosity. For situations where consideration of only elastic cross sections is still justified, we demonstrated that the stopping data imply a significant in-medium cross section reduction, as compared to free, and thus we demonstrate an enhancement of the shear viscosity as compared to that calculated with free cross sections. For reference, shear viscosity calculated with free cross sections and velocities is also given in Fig.~\ref{fig:viscos}.

We now use our newly determined viscosity to explore how close nuclear matter is to being the touted ``perfect fluid''.

\subsection{Lower quantum limit of ratio of viscosity to entropy density}

It has been found theoretically that certain strong coupling limits of gauge theories have a constant ratio of shear viscosity to entropy density regardless of the metric used \cite{danielewicz_dissipative_1985, kovtun_viscosity_2005},
\begin{equation}
 \frac{\eta}{s} = \frac{\hbar}{4 \pi k_\mathrm{B}} \, .
\end{equation}
Moreover, it has been speculated that this value represents a lower limit for all relativistic, finite temperature quantum field theories with zero chemical potential, and for single-component nonrelativistic gases of particles with spin 0 or $1/2$ \cite{kovtun_viscosity_2005}.
We calculate this ratio at intermediate energies to find the proximity of nuclear matter at these energies to this conjectured lower limit.

To find the ratio $\eta/s$, we calculate $\eta$ and $s$ separately.
We simplify finding the entropy density by utilizing the model's ability to describe deuteron yields and use equilibrium conditions relating the ratio of the yield of deuterons and deuteron-like correlations to that of total charge, following the prescriptions of Bertsch and Cugnon \cite{bertsch_entropy_1981} as formulated in Ref.~\cite{csernai_entropy_1986}. We reproduce the formula here:
\begin{equation}
 \sigma = S/A = 3.945 - \ln \left( N_\textrm{d-like} / Z \right) - \frac{1}{8} N_\textrm{d-like} / Z \, ,
\end{equation}
where $N_\textrm{d-like} = N_\textrm{d} + \frac{3}{2}(N_\textrm{t} + N_\textrm{h}) + 2 N_\alpha + \cdots$ and $Z = N_\textrm{p} + N_\textrm{d} + N_\textrm{t} + 2(N_\textrm{h} + N_\alpha) + \cdots$ \cite{bertsch_entropy_1983}.
The bulk of the entropy is produced in regions of hot, dense matter, during the compression and thermalization phase of the reaction. This is also where the stopping signals are generated. Therefore, the density and temperature in that specific space-time region should be used to determine the entropy per volume, $s = \sigma n$, as well as the temperature at which to find the viscosity. In the simulation, we choose a $2\,$fm-radius spherical region centered at the reaction center of mass, during the time of maximal density in that region. The temperature is found assuming that the momentum distribution of the nucleons approximates that of a degenerate relativistic Fermi gas.

\begin{figure}
 \centering
 
 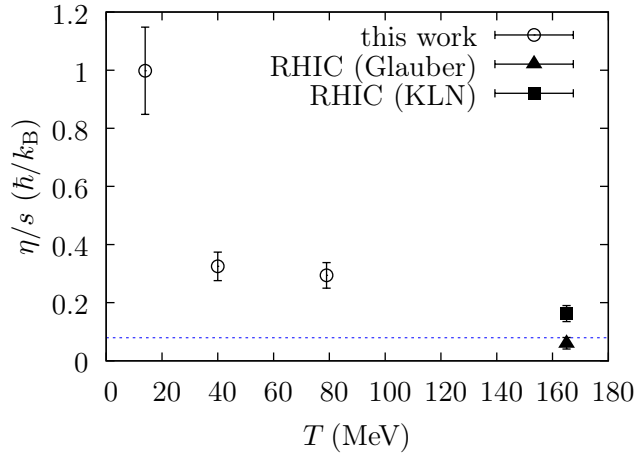
 \caption{This work's estimate for $\eta/s$ values in selected reactions (open circles), alongside the estimates arrived at RHIC energies following the KLN model (square) \cite{adil_eccentricity_2006} and the Glauber model (triangle) \cite{miller_glauber_2007} for the initial conditions. The ratio is given in units of $\hbar/k_\mathrm{B}$. The conjectured lower quantum limit, $1/4\pi$ \cite{kovtun_viscosity_2005}, is shown in a dotted line.}\label{fig:eta-s}
\end{figure}

We choose several representative reactions to find the characteristic temperatures and densities reached at intermediate energies. Listed here in order of decreasing $\eta/s$ and increasing maximal temperature, as represented by open circles in Fig.~\ref{fig:eta-s}, they are \collpt{197}{Au}{197}{Au} at beam energies of 100, 400, and $1000\,$MeV/nucleon, each with a reduced impact parameter $b_\textrm{red} = 1.5$.
Even though this work is at a much lower beam energy, the trend of the nuclear matter looks to match findings at \textsc{RHIC} energies, which use a Monte Carlo Bayesian framework \cite{bernhard_quantifying_2015} following initializations in the Glauber \cite{miller_glauber_2007} and KLN \cite{adil_eccentricity_2006} models.
Indeed, as the temperature approaches the critical temperature for nuclear matter, $\sim 170\,$MeV \cite{karsch_quark_2001}, $\eta/s$ is seen to approach the conjectured lower bound.

\section{Conclusion}

We investigated the viscosity of nuclear matter by adjusting the in-medium nucleon-nucleon cross section to fit nuclear stopping data in terms of several different stopping observables across a wide range of nuclear mass and beam energy.
We found that, for p\textsc{BUU}, an in-medium reduction in the \textsc{NN} cross section is necessary to match a variety of experimental data, and that this need for reduction is consistent across a range of reasonable choices of nuclear mean field.
Using this in-medium nucleon-nucleon cross section, we calculate shear viscosity $\eta$ in nuclear matter, at densities and temperatures representing those encountered in the collisions from which we draw the stopping observables, in a manner consistent with the way we simulate the collisions to match the measured observables.
We argue that the stopping observables better correlate with viscosity than with the details in the cross sections. In calculations of viscosity, the use of reduced cross sections, compared to free-space, increases the viscosity values.
We subsequently calculated the ratio of shear viscosity to entropy per unit volume, $\eta / s$, which is often mentioned in the literature.
We demonstrate that our values for the ratio trend towards that deduced in ultrarelativistic collisions as temperature increases, corresponding to changing beam energy. The calculated ratio is only a few times larger than the speculated absolute lower bound of the ratio.

To benefit from data on stopping at higher energies, where pion production starts to influence the stopping at a significant level, modifications of inelastic processes need to be explored, from which we refrain at present. We do not systematically incorporate the effect of inelastic processes on viscosity at high temperatures either. However, Fig.~\ref{fig:varxz-nocomp-au} suggests a reduction in the rates for inelastic processes in the medium, as compared to free-space extrapolations, and a corresponding increase in viscosity, just as in the case of elastic processes only.

\begin{acknowledgments}
This work was supported by the US National Science Foundation under Grants PHY-1068571 and PHY-1403906.
\end{acknowledgments}

\bibliography{zotero,references-manual}

\begin{thebibliography}{55}%
\makeatletter
\providecommand \@ifxundefined [1]{%
 \@ifx{#1\undefined}
}%
\providecommand \@ifnum [1]{%
 \ifnum #1\expandafter \@firstoftwo
 \else \expandafter \@secondoftwo
 \fi
}%
\providecommand \@ifx [1]{%
 \ifx #1\expandafter \@firstoftwo
 \else \expandafter \@secondoftwo
 \fi
}%
\providecommand \natexlab [1]{#1}%
\providecommand \enquote  [1]{``#1''}%
\providecommand \bibnamefont  [1]{#1}%
\providecommand \bibfnamefont [1]{#1}%
\providecommand \citenamefont [1]{#1}%
\providecommand \href@noop [0]{\@secondoftwo}%
\providecommand \href [0]{\begingroup \@sanitize@url \@href}%
\providecommand \@href[1]{\@@startlink{#1}\@@href}%
\providecommand \@@href[1]{\endgroup#1\@@endlink}%
\providecommand \@sanitize@url [0]{\catcode `\\12\catcode `\$12\catcode
  `\&12\catcode `\#12\catcode `\^12\catcode `\_12\catcode `\%12\relax}%
\providecommand \@@startlink[1]{}%
\providecommand \@@endlink[0]{}%
\providecommand \url  [0]{\begingroup\@sanitize@url \@url }%
\providecommand \@url [1]{\endgroup\@href {#1}{\urlprefix }}%
\providecommand \urlprefix  [0]{URL }%
\providecommand \Eprint [0]{\href }%
\providecommand \doibase [0]{http://dx.doi.org/}%
\providecommand \selectlanguage [0]{\@gobble}%
\providecommand \bibinfo  [0]{\@secondoftwo}%
\providecommand \bibfield  [0]{\@secondoftwo}%
\providecommand \translation [1]{[#1]}%
\providecommand \BibitemOpen [0]{}%
\providecommand \bibitemStop [0]{}%
\providecommand \bibitemNoStop [0]{.\EOS\space}%
\providecommand \EOS [0]{\spacefactor3000\relax}%
\providecommand \BibitemShut  [1]{\csname bibitem#1\endcsname}%
\let\auto@bib@innerbib\@empty
\bibitem [{\citenamefont {Danielewicz}\ and\ \citenamefont
  {Gyulassy}(1985)}]{danielewicz_dissipative_1985}%
  \BibitemOpen
  \bibfield  {author} {\bibinfo {author} {\bibfnamefont {P.}~\bibnamefont
  {Danielewicz}}\ and\ \bibinfo {author} {\bibfnamefont {M.}~\bibnamefont
  {Gyulassy}},\ }\href
  {http://journals.aps.org/prd/abstract/10.1103/PhysRevD.31.53} {\bibfield
  {journal} {\bibinfo  {journal} {Physical Review D}\ }\textbf {\bibinfo
  {volume} {31}},\ \bibinfo {pages} {53} (\bibinfo {year} {1985})}\BibitemShut
  {NoStop}%
\bibitem [{\citenamefont {Kovtun}\ \emph {et~al.}(2005)\citenamefont {Kovtun},
  \citenamefont {Son},\ and\ \citenamefont
  {Starinets}}]{kovtun_viscosity_2005}%
  \BibitemOpen
  \bibfield  {author} {\bibinfo {author} {\bibfnamefont {P.~K.}\ \bibnamefont
  {Kovtun}}, \bibinfo {author} {\bibfnamefont {D.~T.}\ \bibnamefont {Son}}, \
  and\ \bibinfo {author} {\bibfnamefont {A.~O.}\ \bibnamefont {Starinets}},\
  }\href {\doibase 10.1103/PhysRevLett.94.111601} {\bibfield  {journal}
  {\bibinfo  {journal} {Phys. Rev. Lett.}\ }\textbf {\bibinfo {volume} {94}},\
  \bibinfo {pages} {111601} (\bibinfo {year} {2005})}\BibitemShut {NoStop}%
\bibitem [{\citenamefont {Schäfer}(2014)}]{schafer_fluid_2014}%
  \BibitemOpen
  \bibfield  {author} {\bibinfo {author} {\bibfnamefont {T.}~\bibnamefont
  {Schäfer}},\ }\href {\doibase 10.1146/annurev-nucl-102313-025439} {\bibfield
   {journal} {\bibinfo  {journal} {Annual Review of Nuclear and Particle
  Science}\ }\textbf {\bibinfo {volume} {64}},\ \bibinfo {pages} {125}
  (\bibinfo {year} {2014})}\BibitemShut {NoStop}%
\bibitem [{\citenamefont {Lacey}\ \emph {et~al.}(2007)\citenamefont {Lacey},
  \citenamefont {Ajitanand}, \citenamefont {Alexander}, \citenamefont {Chung},
  \citenamefont {Holzmann}, \citenamefont {Issah}, \citenamefont {Taranenko},
  \citenamefont {Danielewicz},\ and\ \citenamefont
  {Stöcker}}]{lacey_has_2007}%
  \BibitemOpen
  \bibfield  {author} {\bibinfo {author} {\bibfnamefont {R.~A.}\ \bibnamefont
  {Lacey}}, \bibinfo {author} {\bibfnamefont {N.~N.}\ \bibnamefont
  {Ajitanand}}, \bibinfo {author} {\bibfnamefont {J.~M.}\ \bibnamefont
  {Alexander}}, \bibinfo {author} {\bibfnamefont {P.}~\bibnamefont {Chung}},
  \bibinfo {author} {\bibfnamefont {W.~G.}\ \bibnamefont {Holzmann}}, \bibinfo
  {author} {\bibfnamefont {M.}~\bibnamefont {Issah}}, \bibinfo {author}
  {\bibfnamefont {A.}~\bibnamefont {Taranenko}}, \bibinfo {author}
  {\bibfnamefont {P.}~\bibnamefont {Danielewicz}}, \ and\ \bibinfo {author}
  {\bibfnamefont {H.}~\bibnamefont {Stöcker}},\ }\href {\doibase
  10.1103/PhysRevLett.98.092301} {\bibfield  {journal} {\bibinfo  {journal}
  {Phys. Rev. Lett.}\ }\textbf {\bibinfo {volume} {98}},\ \bibinfo {pages}
  {092301} (\bibinfo {year} {2007})}\BibitemShut {NoStop}%
\bibitem [{\citenamefont {Li}\ and\ \citenamefont
  {Li}(2017)}]{li_isospin_2017}%
  \BibitemOpen
  \bibfield  {author} {\bibinfo {author} {\bibfnamefont {Q.}~\bibnamefont
  {Li}}\ and\ \bibinfo {author} {\bibfnamefont {Z.}~\bibnamefont {Li}},\ }\href
  {\doibase 10.1016/j.physletb.2017.09.013} {\bibfield  {journal} {\bibinfo
  {journal} {Physics Letters B}\ }\textbf {\bibinfo {volume} {773}},\ \bibinfo
  {pages} {557} (\bibinfo {year} {2017})}\BibitemShut {NoStop}%
\bibitem [{\citenamefont {Danielewicz}\ and\ \citenamefont
  {Bertsch}(1991)}]{danielewicz_production_1991}%
  \BibitemOpen
  \bibfield  {author} {\bibinfo {author} {\bibfnamefont {P.}~\bibnamefont
  {Danielewicz}}\ and\ \bibinfo {author} {\bibfnamefont {G.}~\bibnamefont
  {Bertsch}},\ }\href {\doibase 10.1016/0375-9474(91)90541-D} {\bibfield
  {journal} {\bibinfo  {journal} {Nuclear Physics A}\ }\textbf {\bibinfo
  {volume} {533}},\ \bibinfo {pages} {712} (\bibinfo {year}
  {1991})}\BibitemShut {NoStop}%
\bibitem [{\citenamefont {Danielewicz}\ and\ \citenamefont
  {Pan}(1992)}]{danielewicz_blast_1992}%
  \BibitemOpen
  \bibfield  {author} {\bibinfo {author} {\bibfnamefont {P.}~\bibnamefont
  {Danielewicz}}\ and\ \bibinfo {author} {\bibfnamefont {Q.}~\bibnamefont
  {Pan}},\ }\href {\doibase 10.1103/PhysRevC.46.2002} {\bibfield  {journal}
  {\bibinfo  {journal} {Phys. Rev. C}\ }\textbf {\bibinfo {volume} {46}},\
  \bibinfo {pages} {2002} (\bibinfo {year} {1992})}\BibitemShut {NoStop}%
\bibitem [{\citenamefont {Pan}\ and\ \citenamefont
  {Danielewicz}(1993)}]{pan_sideward_1993}%
  \BibitemOpen
  \bibfield  {author} {\bibinfo {author} {\bibfnamefont {Q.}~\bibnamefont
  {Pan}}\ and\ \bibinfo {author} {\bibfnamefont {P.}~\bibnamefont
  {Danielewicz}},\ }\href {\doibase 10.1103/PhysRevLett.70.2062} {\bibfield
  {journal} {\bibinfo  {journal} {Phys. Rev. Lett.}\ }\textbf {\bibinfo
  {volume} {70}},\ \bibinfo {pages} {2062} (\bibinfo {year}
  {1993})}\BibitemShut {NoStop}%
\bibitem [{\citenamefont {Danielewicz}(2000)}]{danielewicz_determination_2000}%
  \BibitemOpen
  \bibfield  {author} {\bibinfo {author} {\bibfnamefont {P.}~\bibnamefont
  {Danielewicz}},\ }\href {\doibase 10.1016/S0375-9474(00)00083-X} {\bibfield
  {journal} {\bibinfo  {journal} {Nuclear Physics A}\ }\textbf {\bibinfo
  {volume} {673}},\ \bibinfo {pages} {375} (\bibinfo {year}
  {2000})}\BibitemShut {NoStop}%
\bibitem [{\citenamefont {Danielewicz}(2002)}]{danielewicz_hadronic_2002}%
  \BibitemOpen
  \bibfield  {author} {\bibinfo {author} {\bibfnamefont {P.}~\bibnamefont
  {Danielewicz}},\ }\href {http://arxiv.org/abs/nucl-th/0201032} {\bibfield
  {journal} {\bibinfo  {journal} {Acta Phys. Pol. B}\ }\textbf {\bibinfo
  {volume} {33}},\ \bibinfo {pages} {45} (\bibinfo {year} {2002})},\ \bibinfo
  {note} {acta Phys.Polon. B33 (2002) 45-64}\BibitemShut {NoStop}%
\bibitem [{\citenamefont {Barker}(2014)}]{barker_dissipation_2014}%
  \BibitemOpen
  \bibfield  {author} {\bibinfo {author} {\bibfnamefont {B.}~\bibnamefont
  {Barker}},\ }\emph {\bibinfo {title} {Dissipation and dynamics in quantum
  many-body systems}},\ \href@noop {} {\bibinfo {type} {Ph.{D}.}},\ \bibinfo
  {school} {Michigan State University}, \bibinfo {address} {East Lansing, MI,
  USA} (\bibinfo {year} {2014})\BibitemShut {NoStop}%
\bibitem [{\citenamefont {Bertsch}\ and\ \citenamefont
  {Das~Gupta}(1988)}]{bertsch_guide_1988}%
  \BibitemOpen
  \bibfield  {author} {\bibinfo {author} {\bibfnamefont {G.}~\bibnamefont
  {Bertsch}}\ and\ \bibinfo {author} {\bibfnamefont {S.}~\bibnamefont
  {Das~Gupta}},\ }\href {\doibase 10.1016/0370-1573(88)90170-6} {\bibfield
  {journal} {\bibinfo  {journal} {Physics Reports}\ }\textbf {\bibinfo {volume}
  {160}},\ \bibinfo {pages} {189} (\bibinfo {year} {1988})}\BibitemShut
  {NoStop}%
\bibitem [{\citenamefont {Mondal}\ \emph {et~al.}(2017)\citenamefont {Mondal},
  \citenamefont {Pandit}, \citenamefont {Mukhopadhyay}, \citenamefont {Pal},
  \citenamefont {Dey}, \citenamefont {Bhattacharya}, \citenamefont {De},
  \citenamefont {Bhattacharya}, \citenamefont {Bhattacharyya}, \citenamefont
  {Roy}, \citenamefont {Banerjee},\ and\ \citenamefont
  {Banerjee}}]{mondal_experimental_2017}%
  \BibitemOpen
  \bibfield  {author} {\bibinfo {author} {\bibfnamefont {D.}~\bibnamefont
  {Mondal}}, \bibinfo {author} {\bibfnamefont {D.}~\bibnamefont {Pandit}},
  \bibinfo {author} {\bibfnamefont {S.}~\bibnamefont {Mukhopadhyay}}, \bibinfo
  {author} {\bibfnamefont {S.}~\bibnamefont {Pal}}, \bibinfo {author}
  {\bibfnamefont {B.}~\bibnamefont {Dey}}, \bibinfo {author} {\bibfnamefont
  {S.}~\bibnamefont {Bhattacharya}}, \bibinfo {author} {\bibfnamefont
  {A.}~\bibnamefont {De}}, \bibinfo {author} {\bibfnamefont {S.}~\bibnamefont
  {Bhattacharya}}, \bibinfo {author} {\bibfnamefont {S.}~\bibnamefont
  {Bhattacharyya}}, \bibinfo {author} {\bibfnamefont {P.}~\bibnamefont {Roy}},
  \bibinfo {author} {\bibfnamefont {K.}~\bibnamefont {Banerjee}}, \ and\
  \bibinfo {author} {\bibfnamefont {S.~R.}\ \bibnamefont {Banerjee}},\ }\href
  {\doibase 10.1103/PhysRevLett.118.192501} {\bibfield  {journal} {\bibinfo
  {journal} {Physical Review Letters}\ }\textbf {\bibinfo {volume} {118}}
  (\bibinfo {year} {2017}),\ 10.1103/PhysRevLett.118.192501},\ \bibinfo {note}
  {arXiv: 1705.01286}\BibitemShut {NoStop}%
\bibitem [{\citenamefont {Auerbach}\ and\ \citenamefont
  {Shlomo}(2009)}]{auerbach_eta/s_2009}%
  \BibitemOpen
  \bibfield  {author} {\bibinfo {author} {\bibfnamefont {N.}~\bibnamefont
  {Auerbach}}\ and\ \bibinfo {author} {\bibfnamefont {S.}~\bibnamefont
  {Shlomo}},\ }\href {\doibase 10.1103/PhysRevLett.103.172501} {\bibfield
  {journal} {\bibinfo  {journal} {Physical Review Letters}\ }\textbf {\bibinfo
  {volume} {103}},\ \bibinfo {pages} {172501} (\bibinfo {year}
  {2009})}\BibitemShut {NoStop}%
\bibitem [{\citenamefont {Pal}(2010)}]{pal_shear_2010}%
  \BibitemOpen
  \bibfield  {author} {\bibinfo {author} {\bibfnamefont {S.}~\bibnamefont
  {Pal}},\ }\href {\doibase 10.1103/PhysRevC.81.051601} {\bibfield  {journal}
  {\bibinfo  {journal} {Phys. Rev. C}\ }\textbf {\bibinfo {volume} {81}},\
  \bibinfo {pages} {051601} (\bibinfo {year} {2010})}\BibitemShut {NoStop}%
\bibitem [{\citenamefont {Zhou}\ \emph {et~al.}(2012)\citenamefont {Zhou},
  \citenamefont {Ma},\ and\ \citenamefont {Fang}}]{zhou_shear_2012}%
  \BibitemOpen
  \bibfield  {author} {\bibinfo {author} {\bibfnamefont {C.}~\bibnamefont
  {Zhou}}, \bibinfo {author} {\bibfnamefont {Y.}~\bibnamefont {Ma}}, \ and\
  \bibinfo {author} {\bibfnamefont {D.}~\bibnamefont {Fang}},\ }\href {\doibase
  10.1088/1009-0630/14/7/04} {\bibfield  {journal} {\bibinfo  {journal} {Plasma
  Science and Technology}\ }\textbf {\bibinfo {volume} {14}},\ \bibinfo {pages}
  {585} (\bibinfo {year} {2012})}\BibitemShut {NoStop}%
\bibitem [{\citenamefont {Zhou}\ \emph {et~al.}(2014)\citenamefont {Zhou},
  \citenamefont {Ma}, \citenamefont {Fang}, \citenamefont {Zhang},
  \citenamefont {Xu}, \citenamefont {Cao},\ and\ \citenamefont
  {Shen}}]{zhou_correlation_2014}%
  \BibitemOpen
  \bibfield  {author} {\bibinfo {author} {\bibfnamefont {C.~L.}\ \bibnamefont
  {Zhou}}, \bibinfo {author} {\bibfnamefont {Y.~G.}\ \bibnamefont {Ma}},
  \bibinfo {author} {\bibfnamefont {D.~Q.}\ \bibnamefont {Fang}}, \bibinfo
  {author} {\bibfnamefont {G.~Q.}\ \bibnamefont {Zhang}}, \bibinfo {author}
  {\bibfnamefont {J.}~\bibnamefont {Xu}}, \bibinfo {author} {\bibfnamefont
  {X.~G.}\ \bibnamefont {Cao}}, \ and\ \bibinfo {author} {\bibfnamefont
  {W.~Q.}\ \bibnamefont {Shen}},\ }\href {\doibase 10.1103/PhysRevC.90.057601}
  {\bibfield  {journal} {\bibinfo  {journal} {Phys. Rev. C}\ }\textbf {\bibinfo
  {volume} {90}},\ \bibinfo {pages} {057601} (\bibinfo {year}
  {2014})}\BibitemShut {NoStop}%
\bibitem [{\citenamefont {Xu}\ \emph {et~al.}(2013)\citenamefont {Xu},
  \citenamefont {Chen}, \citenamefont {Ko}, \citenamefont {Li},\ and\
  \citenamefont {Ma}}]{xu_shear_2013-1}%
  \BibitemOpen
  \bibfield  {author} {\bibinfo {author} {\bibfnamefont {J.}~\bibnamefont
  {Xu}}, \bibinfo {author} {\bibfnamefont {L.-W.}\ \bibnamefont {Chen}},
  \bibinfo {author} {\bibfnamefont {C.~M.}\ \bibnamefont {Ko}}, \bibinfo
  {author} {\bibfnamefont {B.-A.}\ \bibnamefont {Li}}, \ and\ \bibinfo {author}
  {\bibfnamefont {Y.~G.}\ \bibnamefont {Ma}},\ }\href {\doibase
  10.1016/j.physletb.2013.10.051} {\bibfield  {journal} {\bibinfo  {journal}
  {Physics Letters B}\ }\textbf {\bibinfo {volume} {727}},\ \bibinfo {pages}
  {244} (\bibinfo {year} {2013})}\BibitemShut {NoStop}%
\bibitem [{\citenamefont {Guo}\ \emph {et~al.}(2017)\citenamefont {Guo},
  \citenamefont {Ma}, \citenamefont {He}, \citenamefont {Cao}, \citenamefont
  {Fang}, \citenamefont {Deng},\ and\ \citenamefont
  {Zhou}}]{guo_isovector_2017}%
  \BibitemOpen
  \bibfield  {author} {\bibinfo {author} {\bibfnamefont {C.~Q.}\ \bibnamefont
  {Guo}}, \bibinfo {author} {\bibfnamefont {Y.~G.}\ \bibnamefont {Ma}},
  \bibinfo {author} {\bibfnamefont {W.~B.}\ \bibnamefont {He}}, \bibinfo
  {author} {\bibfnamefont {X.~G.}\ \bibnamefont {Cao}}, \bibinfo {author}
  {\bibfnamefont {D.~Q.}\ \bibnamefont {Fang}}, \bibinfo {author}
  {\bibfnamefont {X.~G.}\ \bibnamefont {Deng}}, \ and\ \bibinfo {author}
  {\bibfnamefont {C.~L.}\ \bibnamefont {Zhou}},\ }\href {\doibase
  10.1103/PhysRevC.95.054622} {\bibfield  {journal} {\bibinfo  {journal} {Phys.
  Rev. C}\ }\textbf {\bibinfo {volume} {95}},\ \bibinfo {pages} {054622}
  (\bibinfo {year} {2017})}\BibitemShut {NoStop}%
\bibitem [{\citenamefont {Danielewicz}(1984)}]{danielewicz_transport_1984}%
  \BibitemOpen
  \bibfield  {author} {\bibinfo {author} {\bibfnamefont {P.}~\bibnamefont
  {Danielewicz}},\ }\href {\doibase 10.1016/0370-2693(84)91010-4} {\bibfield
  {journal} {\bibinfo  {journal} {Physics Letters B}\ }\textbf {\bibinfo
  {volume} {146}},\ \bibinfo {pages} {168} (\bibinfo {year}
  {1984})}\BibitemShut {NoStop}%
\bibitem [{\citenamefont {Shi}\ and\ \citenamefont
  {Danielewicz}(2003)}]{shi_nuclear_2003}%
  \BibitemOpen
  \bibfield  {author} {\bibinfo {author} {\bibfnamefont {L.}~\bibnamefont
  {Shi}}\ and\ \bibinfo {author} {\bibfnamefont {P.}~\bibnamefont
  {Danielewicz}},\ }\href {\doibase 10.1103/PhysRevC.68.064604} {\bibfield
  {journal} {\bibinfo  {journal} {Phys. Rev. C}\ }\textbf {\bibinfo {volume}
  {68}},\ \bibinfo {pages} {064604} (\bibinfo {year} {2003})}\BibitemShut
  {NoStop}%
\bibitem [{\citenamefont {Pandharipande}\ and\ \citenamefont
  {Pieper}(1992)}]{pandharipande_nuclear_1992}%
  \BibitemOpen
  \bibfield  {author} {\bibinfo {author} {\bibfnamefont {V.~R.}\ \bibnamefont
  {Pandharipande}}\ and\ \bibinfo {author} {\bibfnamefont {S.~C.}\ \bibnamefont
  {Pieper}},\ }\href {\doibase 10.1103/PhysRevC.45.791} {\bibfield  {journal}
  {\bibinfo  {journal} {Phys. Rev. C}\ }\textbf {\bibinfo {volume} {45}},\
  \bibinfo {pages} {791} (\bibinfo {year} {1992})}\BibitemShut {NoStop}%
\bibitem [{\citenamefont {Persram}\ and\ \citenamefont
  {Gale}(2002)}]{persram_elliptic_2002}%
  \BibitemOpen
  \bibfield  {author} {\bibinfo {author} {\bibfnamefont {D.}~\bibnamefont
  {Persram}}\ and\ \bibinfo {author} {\bibfnamefont {C.}~\bibnamefont {Gale}},\
  }\href {\doibase 10.1103/PhysRevC.65.064611} {\bibfield  {journal} {\bibinfo
  {journal} {Phys. Rev. C}\ }\textbf {\bibinfo {volume} {65}},\ \bibinfo
  {pages} {064611} (\bibinfo {year} {2002})}\BibitemShut {NoStop}%
\bibitem [{\citenamefont {Li}\ and\ \citenamefont
  {Chen}(2005)}]{li_nucleon-nucleon_2005}%
  \BibitemOpen
  \bibfield  {author} {\bibinfo {author} {\bibfnamefont {B.-A.}\ \bibnamefont
  {Li}}\ and\ \bibinfo {author} {\bibfnamefont {L.-W.}\ \bibnamefont {Chen}},\
  }\href {\doibase 10.1103/PhysRevC.72.064611} {\bibfield  {journal} {\bibinfo
  {journal} {Phys. Rev. C}\ }\textbf {\bibinfo {volume} {72}},\ \bibinfo
  {pages} {064611} (\bibinfo {year} {2005})}\BibitemShut {NoStop}%
\bibitem [{\citenamefont {Sammarruca}(2014)}]{sammarruca_microscopic_2014}%
  \BibitemOpen
  \bibfield  {author} {\bibinfo {author} {\bibfnamefont {F.}~\bibnamefont
  {Sammarruca}},\ }\href@noop {} {\bibfield  {journal} {\bibinfo  {journal}
  {Eur. Phys. J. A}\ }\textbf {\bibinfo {volume} {50}},\ \bibinfo {pages} {22}
  (\bibinfo {year} {2014})}\BibitemShut {NoStop}%
\bibitem [{\citenamefont {Yanhuang}\ and\ \citenamefont
  {Di~Toro}(1989)}]{cai_yanhuang_semiclassical_1989}%
  \BibitemOpen
  \bibfield  {author} {\bibinfo {author} {\bibfnamefont {C.}~\bibnamefont
  {Yanhuang}}\ and\ \bibinfo {author} {\bibfnamefont {M.}~\bibnamefont
  {Di~Toro}},\ }\href {\doibase 10.1103/PhysRevC.39.105} {\bibfield  {journal}
  {\bibinfo  {journal} {Phys. Rev. C}\ }\textbf {\bibinfo {volume} {39}},\
  \bibinfo {pages} {105} (\bibinfo {year} {1989})}\BibitemShut {NoStop}%
\bibitem [{\citenamefont {Gaitanos}\ \emph {et~al.}(2004)\citenamefont
  {Gaitanos}, \citenamefont {Colonna}, \citenamefont {Di~Toro},\ and\
  \citenamefont {Wolter}}]{gaitanos_stopping_2004}%
  \BibitemOpen
  \bibfield  {author} {\bibinfo {author} {\bibfnamefont {T.}~\bibnamefont
  {Gaitanos}}, \bibinfo {author} {\bibfnamefont {M.}~\bibnamefont {Colonna}},
  \bibinfo {author} {\bibfnamefont {M.}~\bibnamefont {Di~Toro}}, \ and\
  \bibinfo {author} {\bibfnamefont {H.}~\bibnamefont {Wolter}},\ }\href
  {\doibase 10.1016/j.physletb.2004.05.080} {\bibfield  {journal} {\bibinfo
  {journal} {Physics Letters B}\ }\textbf {\bibinfo {volume} {595}},\ \bibinfo
  {pages} {209} (\bibinfo {year} {2004})}\BibitemShut {NoStop}%
\bibitem [{\citenamefont {Zhou}\ \emph {et~al.}(2013)\citenamefont {Zhou},
  \citenamefont {Ma}, \citenamefont {Fang},\ and\ \citenamefont
  {Zhang}}]{zhou_thermodynamic_2013}%
  \BibitemOpen
  \bibfield  {author} {\bibinfo {author} {\bibfnamefont {C.~L.}\ \bibnamefont
  {Zhou}}, \bibinfo {author} {\bibfnamefont {Y.~G.}\ \bibnamefont {Ma}},
  \bibinfo {author} {\bibfnamefont {D.~Q.}\ \bibnamefont {Fang}}, \ and\
  \bibinfo {author} {\bibfnamefont {G.~Q.}\ \bibnamefont {Zhang}},\ }\href
  {\doibase 10.1103/PhysRevC.88.024604} {\bibfield  {journal} {\bibinfo
  {journal} {Phys. Rev. C}\ }\textbf {\bibinfo {volume} {88}},\ \bibinfo
  {pages} {024604} (\bibinfo {year} {2013})}\BibitemShut {NoStop}%
\bibitem [{\citenamefont {Basrak}\ \emph {et~al.}(2016)\citenamefont {Basrak},
  \citenamefont {Eudes},\ and\ \citenamefont {de~la
  Mota}}]{basrak_aspects_2016}%
  \BibitemOpen
  \bibfield  {author} {\bibinfo {author} {\bibfnamefont {Z.}~\bibnamefont
  {Basrak}}, \bibinfo {author} {\bibfnamefont {P.}~\bibnamefont {Eudes}}, \
  and\ \bibinfo {author} {\bibfnamefont {V.}~\bibnamefont {de~la Mota}},\
  }\href {\doibase 10.1103/PhysRevC.93.054609} {\bibfield  {journal} {\bibinfo
  {journal} {Phys. Rev. C}\ }\textbf {\bibinfo {volume} {93}},\ \bibinfo
  {pages} {054609} (\bibinfo {year} {2016})}\BibitemShut {NoStop}%
\bibitem [{\citenamefont {Alm}\ \emph {et~al.}(1994{\natexlab{a}})\citenamefont
  {Alm}, \citenamefont {Röpke},\ and\ \citenamefont
  {Schmidt}}]{alm_critical_1994}%
  \BibitemOpen
  \bibfield  {author} {\bibinfo {author} {\bibfnamefont {T.}~\bibnamefont
  {Alm}}, \bibinfo {author} {\bibfnamefont {G.}~\bibnamefont {Röpke}}, \ and\
  \bibinfo {author} {\bibfnamefont {M.}~\bibnamefont {Schmidt}},\ }\href
  {\doibase 10.1103/PhysRevC.50.31} {\bibfield  {journal} {\bibinfo  {journal}
  {Phys. Rev. C}\ }\textbf {\bibinfo {volume} {50}},\ \bibinfo {pages} {31}
  (\bibinfo {year} {1994}{\natexlab{a}})}\BibitemShut {NoStop}%
\bibitem [{\citenamefont {Gaitanos}\ \emph {et~al.}(2005)\citenamefont
  {Gaitanos}, \citenamefont {Fuchs},\ and\ \citenamefont
  {Wolter}}]{gaitanos_nuclear_2005}%
  \BibitemOpen
  \bibfield  {author} {\bibinfo {author} {\bibfnamefont {T.}~\bibnamefont
  {Gaitanos}}, \bibinfo {author} {\bibfnamefont {C.}~\bibnamefont {Fuchs}}, \
  and\ \bibinfo {author} {\bibfnamefont {H.}~\bibnamefont {Wolter}},\ }\href
  {\doibase 10.1016/j.physletb.2005.01.069} {\bibfield  {journal} {\bibinfo
  {journal} {Physics Letters B}\ }\textbf {\bibinfo {volume} {609}},\ \bibinfo
  {pages} {241} (\bibinfo {year} {2005})}\BibitemShut {NoStop}%
\bibitem [{\citenamefont {Westfall}\ \emph {et~al.}(1993)\citenamefont
  {Westfall}, \citenamefont {Bauer}, \citenamefont {Craig}, \citenamefont
  {Cronqvist}, \citenamefont {Gaultieri}, \citenamefont {Hannuschke},
  \citenamefont {Klakow}, \citenamefont {Li}, \citenamefont {Reposeur},
  \citenamefont {Vander~Molen}, \citenamefont {Wilson}, \citenamefont
  {Winfield}, \citenamefont {Yee}, \citenamefont {Yennello}, \citenamefont
  {Lacey}, \citenamefont {Elmaani}, \citenamefont {Lauret}, \citenamefont
  {Nadasen},\ and\ \citenamefont {Norbeck}}]{westfall_mass_1993}%
  \BibitemOpen
  \bibfield  {author} {\bibinfo {author} {\bibfnamefont {G.~D.}\ \bibnamefont
  {Westfall}}, \bibinfo {author} {\bibfnamefont {W.}~\bibnamefont {Bauer}},
  \bibinfo {author} {\bibfnamefont {D.}~\bibnamefont {Craig}}, \bibinfo
  {author} {\bibfnamefont {M.}~\bibnamefont {Cronqvist}}, \bibinfo {author}
  {\bibfnamefont {E.}~\bibnamefont {Gaultieri}}, \bibinfo {author}
  {\bibfnamefont {S.}~\bibnamefont {Hannuschke}}, \bibinfo {author}
  {\bibfnamefont {D.}~\bibnamefont {Klakow}}, \bibinfo {author} {\bibfnamefont
  {T.}~\bibnamefont {Li}}, \bibinfo {author} {\bibfnamefont {T.}~\bibnamefont
  {Reposeur}}, \bibinfo {author} {\bibfnamefont {A.~M.}\ \bibnamefont
  {Vander~Molen}}, \bibinfo {author} {\bibfnamefont {W.~K.}\ \bibnamefont
  {Wilson}}, \bibinfo {author} {\bibfnamefont {J.~S.}\ \bibnamefont
  {Winfield}}, \bibinfo {author} {\bibfnamefont {J.}~\bibnamefont {Yee}},
  \bibinfo {author} {\bibfnamefont {S.~J.}\ \bibnamefont {Yennello}}, \bibinfo
  {author} {\bibfnamefont {R.}~\bibnamefont {Lacey}}, \bibinfo {author}
  {\bibfnamefont {A.}~\bibnamefont {Elmaani}}, \bibinfo {author} {\bibfnamefont
  {J.}~\bibnamefont {Lauret}}, \bibinfo {author} {\bibfnamefont
  {A.}~\bibnamefont {Nadasen}}, \ and\ \bibinfo {author} {\bibfnamefont
  {E.}~\bibnamefont {Norbeck}},\ }\href {\doibase 10.1103/PhysRevLett.71.1986}
  {\bibfield  {journal} {\bibinfo  {journal} {Phys. Rev. Lett.}\ }\textbf
  {\bibinfo {volume} {71}},\ \bibinfo {pages} {1986} (\bibinfo {year}
  {1993})}\BibitemShut {NoStop}%
\bibitem [{\citenamefont {Alm}\ \emph {et~al.}(1994{\natexlab{b}})\citenamefont
  {Alm}, \citenamefont {R\"opke},\ and\ \citenamefont
  {Schmidt}}]{my_alm_critical_1994}%
  \BibitemOpen
  \bibfield  {author} {\bibinfo {author} {\bibfnamefont {T.}~\bibnamefont
  {Alm}}, \bibinfo {author} {\bibfnamefont {G.}~\bibnamefont {R\"opke}}, \ and\
  \bibinfo {author} {\bibfnamefont {M.}~\bibnamefont {Schmidt}},\ }\href
  {\doibase 10.1103/PhysRevC.50.31} {\bibfield  {journal} {\bibinfo  {journal}
  {Physical Review C}\ }\textbf {\bibinfo {volume} {50}},\ \bibinfo {pages}
  {31} (\bibinfo {year} {1994}{\natexlab{b}})}\BibitemShut {NoStop}%
\bibitem [{\citenamefont {Alm}\ \emph {et~al.}(1995)\citenamefont {Alm},
  \citenamefont {Röpke}, \citenamefont {Bauer}, \citenamefont {Daffin},\ and\
  \citenamefont {Schmidt}}]{alm_-medium_1995}%
  \BibitemOpen
  \bibfield  {author} {\bibinfo {author} {\bibfnamefont {T.}~\bibnamefont
  {Alm}}, \bibinfo {author} {\bibfnamefont {G.}~\bibnamefont {Röpke}},
  \bibinfo {author} {\bibfnamefont {W.}~\bibnamefont {Bauer}}, \bibinfo
  {author} {\bibfnamefont {F.}~\bibnamefont {Daffin}}, \ and\ \bibinfo {author}
  {\bibfnamefont {M.}~\bibnamefont {Schmidt}},\ }\href {\doibase
  10.1016/0375-9474(95)00026-W} {\bibfield  {journal} {\bibinfo  {journal}
  {Nuclear Physics A}\ }\textbf {\bibinfo {volume} {587}},\ \bibinfo {pages}
  {815} (\bibinfo {year} {1995})}\BibitemShut {NoStop}%
\bibitem [{\citenamefont {Fuchs}\ \emph {et~al.}(2001)\citenamefont {Fuchs},
  \citenamefont {Faessler},\ and\ \citenamefont
  {El-Shabshiry}}]{fuchs_off-shell_2001}%
  \BibitemOpen
  \bibfield  {author} {\bibinfo {author} {\bibfnamefont {C.}~\bibnamefont
  {Fuchs}}, \bibinfo {author} {\bibfnamefont {A.}~\bibnamefont {Faessler}}, \
  and\ \bibinfo {author} {\bibfnamefont {M.}~\bibnamefont {El-Shabshiry}},\
  }\href {\doibase 10.1103/PhysRevC.64.024003} {\bibfield  {journal} {\bibinfo
  {journal} {Phys. Rev. C}\ }\textbf {\bibinfo {volume} {64}},\ \bibinfo
  {pages} {024003} (\bibinfo {year} {2001})}\BibitemShut {NoStop}%
\bibitem [{\citenamefont {Nicholson}\ and\ \citenamefont
  {Halpern}(1959)}]{nicholson_direct-interaction_1959}%
  \BibitemOpen
  \bibfield  {author} {\bibinfo {author} {\bibfnamefont {W.~J.}\ \bibnamefont
  {Nicholson}}\ and\ \bibinfo {author} {\bibfnamefont {I.}~\bibnamefont
  {Halpern}},\ }\href {\doibase 10.1103/PhysRev.116.175} {\bibfield  {journal}
  {\bibinfo  {journal} {Phys. Rev.}\ }\textbf {\bibinfo {volume} {116}},\
  \bibinfo {pages} {175} (\bibinfo {year} {1959})}\BibitemShut {NoStop}%
\bibitem [{\citenamefont {Sikkeland}\ \emph {et~al.}(1962)\citenamefont
  {Sikkeland}, \citenamefont {Haines},\ and\ \citenamefont
  {Viola}}]{sikkeland_momentum_1962}%
  \BibitemOpen
  \bibfield  {author} {\bibinfo {author} {\bibfnamefont {T.}~\bibnamefont
  {Sikkeland}}, \bibinfo {author} {\bibfnamefont {E.~L.}\ \bibnamefont
  {Haines}}, \ and\ \bibinfo {author} {\bibfnamefont {V.~E.}\ \bibnamefont
  {Viola}},\ }\href {\doibase 10.1103/PhysRev.125.1350} {\bibfield  {journal}
  {\bibinfo  {journal} {Phys. Rev.}\ }\textbf {\bibinfo {volume} {125}},\
  \bibinfo {pages} {1350} (\bibinfo {year} {1962})}\BibitemShut {NoStop}%
\bibitem [{\citenamefont {Viola}\ \emph {et~al.}(1982)\citenamefont {Viola},
  \citenamefont {Back}, \citenamefont {Wolf}, \citenamefont {Awes},
  \citenamefont {Gelbke},\ and\ \citenamefont {Breuer}}]{viola_linear_1982}%
  \BibitemOpen
  \bibfield  {author} {\bibinfo {author} {\bibfnamefont {V.~E.}\ \bibnamefont
  {Viola}}, \bibinfo {author} {\bibfnamefont {B.~B.}\ \bibnamefont {Back}},
  \bibinfo {author} {\bibfnamefont {K.~L.}\ \bibnamefont {Wolf}}, \bibinfo
  {author} {\bibfnamefont {T.~C.}\ \bibnamefont {Awes}}, \bibinfo {author}
  {\bibfnamefont {C.~K.}\ \bibnamefont {Gelbke}}, \ and\ \bibinfo {author}
  {\bibfnamefont {H.}~\bibnamefont {Breuer}},\ }\href {\doibase
  10.1103/PhysRevC.26.178} {\bibfield  {journal} {\bibinfo  {journal} {Phys.
  Rev. C}\ }\textbf {\bibinfo {volume} {26}},\ \bibinfo {pages} {178} (\bibinfo
  {year} {1982})}\BibitemShut {NoStop}%
\bibitem [{\citenamefont {Colin}\ \emph
  {et~al.}(1998{\natexlab{a}})\citenamefont {Colin}, \citenamefont {Sun},
  \citenamefont {Ajitanand}, \citenamefont {Alexander}, \citenamefont {Barton},
  \citenamefont {DeYoung}, \citenamefont {Elmaani}, \citenamefont {Gelderloos},
  \citenamefont {Gualtieri}, \citenamefont {Guinet}, \citenamefont
  {Hannuschke}, \citenamefont {Jasma}, \citenamefont {Kowalski}, \citenamefont
  {Lacey}, \citenamefont {Lauret}, \citenamefont {Norbeck}, \citenamefont
  {Pak}, \citenamefont {Peaslee}, \citenamefont {Stern}, \citenamefont {Stone},
  \citenamefont {Sundbeck}, \citenamefont {Vander~Molen}, \citenamefont
  {Westfall},\ and\ \citenamefont {Yee}}]{colin_splintering_1998}%
  \BibitemOpen
  \bibfield  {author} {\bibinfo {author} {\bibfnamefont {E.}~\bibnamefont
  {Colin}}, \bibinfo {author} {\bibfnamefont {R.}~\bibnamefont {Sun}}, \bibinfo
  {author} {\bibfnamefont {N.~N.}\ \bibnamefont {Ajitanand}}, \bibinfo {author}
  {\bibfnamefont {J.~M.}\ \bibnamefont {Alexander}}, \bibinfo {author}
  {\bibfnamefont {M.~A.}\ \bibnamefont {Barton}}, \bibinfo {author}
  {\bibfnamefont {P.~A.}\ \bibnamefont {DeYoung}}, \bibinfo {author}
  {\bibfnamefont {A.}~\bibnamefont {Elmaani}}, \bibinfo {author} {\bibfnamefont
  {C.~J.}\ \bibnamefont {Gelderloos}}, \bibinfo {author} {\bibfnamefont
  {E.~E.}\ \bibnamefont {Gualtieri}}, \bibinfo {author} {\bibfnamefont
  {D.}~\bibnamefont {Guinet}}, \bibinfo {author} {\bibfnamefont
  {S.}~\bibnamefont {Hannuschke}}, \bibinfo {author} {\bibfnamefont {J.~A.}\
  \bibnamefont {Jasma}}, \bibinfo {author} {\bibfnamefont {L.}~\bibnamefont
  {Kowalski}}, \bibinfo {author} {\bibfnamefont {R.~A.}\ \bibnamefont {Lacey}},
  \bibinfo {author} {\bibfnamefont {J.}~\bibnamefont {Lauret}}, \bibinfo
  {author} {\bibfnamefont {E.}~\bibnamefont {Norbeck}}, \bibinfo {author}
  {\bibfnamefont {R.}~\bibnamefont {Pak}}, \bibinfo {author} {\bibfnamefont
  {G.~F.}\ \bibnamefont {Peaslee}}, \bibinfo {author} {\bibfnamefont
  {M.}~\bibnamefont {Stern}}, \bibinfo {author} {\bibfnamefont {N.~T.~B.}\
  \bibnamefont {Stone}}, \bibinfo {author} {\bibfnamefont {S.~D.}\ \bibnamefont
  {Sundbeck}}, \bibinfo {author} {\bibfnamefont {A.~M.}\ \bibnamefont
  {Vander~Molen}}, \bibinfo {author} {\bibfnamefont {G.~D.}\ \bibnamefont
  {Westfall}}, \ and\ \bibinfo {author} {\bibfnamefont {J.}~\bibnamefont
  {Yee}},\ }\href {\doibase 10.1103/PhysRevC.57.R1032} {\bibfield  {journal}
  {\bibinfo  {journal} {Phys. Rev. C}\ }\textbf {\bibinfo {volume} {57}},\
  \bibinfo {pages} {R1032} (\bibinfo {year} {1998}{\natexlab{a}})}\BibitemShut
  {NoStop}%
\bibitem [{\citenamefont {Colin}\ \emph
  {et~al.}(1998{\natexlab{b}})\citenamefont {Colin}, \citenamefont {Sun},
  \citenamefont {Ajitanand}, \citenamefont {Alexander}, \citenamefont {Barton},
  \citenamefont {{DeYoung}}, \citenamefont {Elmaani}, \citenamefont
  {Gelderloos}, \citenamefont {Gualtieri}, \citenamefont {Guinet},
  \citenamefont {Hannuschke}, \citenamefont {Jasma}, \citenamefont {Kowalski},
  \citenamefont {Lacey}, \citenamefont {Lauret}, \citenamefont {Norbeck},
  \citenamefont {Pak}, \citenamefont {Peaslee}, \citenamefont {Stern},
  \citenamefont {Stone}, \citenamefont {Sundbeck}, \citenamefont
  {Vander~Molen}, \citenamefont {Westfall},\ and\ \citenamefont
  {Yee}}]{my_colin_splintering_1998}%
  \BibitemOpen
  \bibfield  {author} {\bibinfo {author} {\bibfnamefont {E.}~\bibnamefont
  {Colin}}, \bibinfo {author} {\bibfnamefont {R.}~\bibnamefont {Sun}}, \bibinfo
  {author} {\bibfnamefont {N.~N.}\ \bibnamefont {Ajitanand}}, \bibinfo {author}
  {\bibfnamefont {J.~M.}\ \bibnamefont {Alexander}}, \bibinfo {author}
  {\bibfnamefont {M.~A.}\ \bibnamefont {Barton}}, \bibinfo {author}
  {\bibfnamefont {P.~A.}\ \bibnamefont {{DeYoung}}}, \bibinfo {author}
  {\bibfnamefont {A.}~\bibnamefont {Elmaani}}, \bibinfo {author} {\bibfnamefont
  {C.~J.}\ \bibnamefont {Gelderloos}}, \bibinfo {author} {\bibfnamefont
  {E.~E.}\ \bibnamefont {Gualtieri}}, \bibinfo {author} {\bibfnamefont
  {D.}~\bibnamefont {Guinet}}, \bibinfo {author} {\bibfnamefont
  {S.}~\bibnamefont {Hannuschke}}, \bibinfo {author} {\bibfnamefont {J.~A.}\
  \bibnamefont {Jasma}}, \bibinfo {author} {\bibfnamefont {L.}~\bibnamefont
  {Kowalski}}, \bibinfo {author} {\bibfnamefont {R.~A.}\ \bibnamefont {Lacey}},
  \bibinfo {author} {\bibfnamefont {J.}~\bibnamefont {Lauret}}, \bibinfo
  {author} {\bibfnamefont {E.}~\bibnamefont {Norbeck}}, \bibinfo {author}
  {\bibfnamefont {R.}~\bibnamefont {Pak}}, \bibinfo {author} {\bibfnamefont
  {G.~F.}\ \bibnamefont {Peaslee}}, \bibinfo {author} {\bibfnamefont
  {M.}~\bibnamefont {Stern}}, \bibinfo {author} {\bibfnamefont {N.~T.~B.}\
  \bibnamefont {Stone}}, \bibinfo {author} {\bibfnamefont {S.~D.}\ \bibnamefont
  {Sundbeck}}, \bibinfo {author} {\bibfnamefont {A.~M.}\ \bibnamefont
  {Vander~Molen}}, \bibinfo {author} {\bibfnamefont {G.~D.}\ \bibnamefont
  {Westfall}}, \ and\ \bibinfo {author} {\bibfnamefont {J.}~\bibnamefont
  {Yee}},\ }\href {\doibase 10.1103/PhysRevC.57.R1032} {\bibfield  {journal}
  {\bibinfo  {journal} {Physical Review C}\ }\textbf {\bibinfo {volume} {57}},\
  \bibinfo {pages} {R1032} (\bibinfo {year} {1998}{\natexlab{b}})}\BibitemShut
  {NoStop}%
\bibitem [{\citenamefont {Reisdorf}\ \emph {et~al.}(2007)\citenamefont
  {Reisdorf}, \citenamefont {Stockmeier}, \citenamefont {Andronic},
  \citenamefont {Benabderrahmane}, \citenamefont {Hartmann}, \citenamefont
  {Herrmann}, \citenamefont {Hildenbrand}, \citenamefont {Kim}, \citenamefont
  {Kiš}, \citenamefont {Koczoń}, \citenamefont {Kress}, \citenamefont
  {Leifels}, \citenamefont {Lopez}, \citenamefont {Merschmeyer}, \citenamefont
  {Schüttauf}, \citenamefont {Barret}, \citenamefont {Basrak}, \citenamefont
  {Bastid}, \citenamefont {Čaplar}, \citenamefont {Crochet}, \citenamefont
  {Dupieux}, \citenamefont {Dželalija}, \citenamefont {Fodor}, \citenamefont
  {Grishkin}, \citenamefont {Hong}, \citenamefont {Kang}, \citenamefont
  {Kecskemeti}, \citenamefont {Kirejczyk}, \citenamefont {Korolija},
  \citenamefont {Kotte}, \citenamefont {Lebedev}, \citenamefont {Matulewicz},
  \citenamefont {Neubert}, \citenamefont {Petrovici}, \citenamefont {Rami},
  \citenamefont {Ryu}, \citenamefont {Seres}, \citenamefont {Sikora},
  \citenamefont {Sim}, \citenamefont {Simion}, \citenamefont
  {Siwek-Wilczyńska}, \citenamefont {Smolyankin}, \citenamefont {Stoicea},
  \citenamefont {Tymiński}, \citenamefont {Wiśniewski}, \citenamefont
  {Wohlfarth}, \citenamefont {Xiao}, \citenamefont {Xu}, \citenamefont
  {Yushmanov},\ and\ \citenamefont {Zhilin}}]{reisdorf_systematics_2007}%
  \BibitemOpen
  \bibfield  {author} {\bibinfo {author} {\bibfnamefont {W.}~\bibnamefont
  {Reisdorf}}, \bibinfo {author} {\bibfnamefont {M.}~\bibnamefont
  {Stockmeier}}, \bibinfo {author} {\bibfnamefont {A.}~\bibnamefont
  {Andronic}}, \bibinfo {author} {\bibfnamefont {M.}~\bibnamefont
  {Benabderrahmane}}, \bibinfo {author} {\bibfnamefont {O.}~\bibnamefont
  {Hartmann}}, \bibinfo {author} {\bibfnamefont {N.}~\bibnamefont {Herrmann}},
  \bibinfo {author} {\bibfnamefont {K.}~\bibnamefont {Hildenbrand}}, \bibinfo
  {author} {\bibfnamefont {Y.}~\bibnamefont {Kim}}, \bibinfo {author}
  {\bibfnamefont {M.}~\bibnamefont {Kiš}}, \bibinfo {author} {\bibfnamefont
  {P.}~\bibnamefont {Koczoń}}, \bibinfo {author} {\bibfnamefont
  {T.}~\bibnamefont {Kress}}, \bibinfo {author} {\bibfnamefont
  {Y.}~\bibnamefont {Leifels}}, \bibinfo {author} {\bibfnamefont
  {X.}~\bibnamefont {Lopez}}, \bibinfo {author} {\bibfnamefont
  {M.}~\bibnamefont {Merschmeyer}}, \bibinfo {author} {\bibfnamefont
  {A.}~\bibnamefont {Schüttauf}}, \bibinfo {author} {\bibfnamefont
  {V.}~\bibnamefont {Barret}}, \bibinfo {author} {\bibfnamefont
  {Z.}~\bibnamefont {Basrak}}, \bibinfo {author} {\bibfnamefont
  {N.}~\bibnamefont {Bastid}}, \bibinfo {author} {\bibfnamefont
  {R.}~\bibnamefont {Čaplar}}, \bibinfo {author} {\bibfnamefont
  {P.}~\bibnamefont {Crochet}}, \bibinfo {author} {\bibfnamefont
  {P.}~\bibnamefont {Dupieux}}, \bibinfo {author} {\bibfnamefont
  {M.}~\bibnamefont {Dželalija}}, \bibinfo {author} {\bibfnamefont
  {Z.}~\bibnamefont {Fodor}}, \bibinfo {author} {\bibfnamefont
  {Y.}~\bibnamefont {Grishkin}}, \bibinfo {author} {\bibfnamefont
  {B.}~\bibnamefont {Hong}}, \bibinfo {author} {\bibfnamefont {T.}~\bibnamefont
  {Kang}}, \bibinfo {author} {\bibfnamefont {J.}~\bibnamefont {Kecskemeti}},
  \bibinfo {author} {\bibfnamefont {M.}~\bibnamefont {Kirejczyk}}, \bibinfo
  {author} {\bibfnamefont {M.}~\bibnamefont {Korolija}}, \bibinfo {author}
  {\bibfnamefont {R.}~\bibnamefont {Kotte}}, \bibinfo {author} {\bibfnamefont
  {A.}~\bibnamefont {Lebedev}}, \bibinfo {author} {\bibfnamefont
  {T.}~\bibnamefont {Matulewicz}}, \bibinfo {author} {\bibfnamefont
  {W.}~\bibnamefont {Neubert}}, \bibinfo {author} {\bibfnamefont
  {M.}~\bibnamefont {Petrovici}}, \bibinfo {author} {\bibfnamefont
  {F.}~\bibnamefont {Rami}}, \bibinfo {author} {\bibfnamefont {M.}~\bibnamefont
  {Ryu}}, \bibinfo {author} {\bibfnamefont {Z.}~\bibnamefont {Seres}}, \bibinfo
  {author} {\bibfnamefont {B.}~\bibnamefont {Sikora}}, \bibinfo {author}
  {\bibfnamefont {K.}~\bibnamefont {Sim}}, \bibinfo {author} {\bibfnamefont
  {V.}~\bibnamefont {Simion}}, \bibinfo {author} {\bibfnamefont
  {K.}~\bibnamefont {Siwek-Wilczyńska}}, \bibinfo {author} {\bibfnamefont
  {V.}~\bibnamefont {Smolyankin}}, \bibinfo {author} {\bibfnamefont
  {G.}~\bibnamefont {Stoicea}}, \bibinfo {author} {\bibfnamefont
  {Z.}~\bibnamefont {Tymiński}}, \bibinfo {author} {\bibfnamefont
  {K.}~\bibnamefont {Wiśniewski}}, \bibinfo {author} {\bibfnamefont
  {D.}~\bibnamefont {Wohlfarth}}, \bibinfo {author} {\bibfnamefont
  {Z.}~\bibnamefont {Xiao}}, \bibinfo {author} {\bibfnamefont {H.}~\bibnamefont
  {Xu}}, \bibinfo {author} {\bibfnamefont {I.}~\bibnamefont {Yushmanov}}, \
  and\ \bibinfo {author} {\bibfnamefont {A.}~\bibnamefont {Zhilin}},\ }\href
  {\doibase 10.1016/j.nuclphysa.2006.10.085} {\bibfield  {journal} {\bibinfo
  {journal} {Nuclear Physics A}\ }\textbf {\bibinfo {volume} {781}},\ \bibinfo
  {pages} {459} (\bibinfo {year} {2007})}\BibitemShut {NoStop}%
\bibitem [{\citenamefont {Reisdorf}\ \emph
  {et~al.}(2010{\natexlab{a}})\citenamefont {Reisdorf}, \citenamefont
  {Andronic}, \citenamefont {Averbeck}, \citenamefont {Benabderrahmane},
  \citenamefont {Hartmann}, \citenamefont {Herrmann}, \citenamefont
  {Hildenbrand}, \citenamefont {Kang}, \citenamefont {Kim}, \citenamefont
  {Kiš}, \citenamefont {Koczoń}, \citenamefont {Kress}, \citenamefont
  {Leifels}, \citenamefont {Merschmeyer}, \citenamefont {Piasecki},
  \citenamefont {Schüttauf}, \citenamefont {Stockmeier}, \citenamefont
  {Barret}, \citenamefont {Basrak}, \citenamefont {Bastid}, \citenamefont
  {Čaplar}, \citenamefont {Crochet}, \citenamefont {Dupieux}, \citenamefont
  {Dželalija}, \citenamefont {Fodor}, \citenamefont {Gasik}, \citenamefont
  {Grishkin}, \citenamefont {Hong}, \citenamefont {Kecskemeti}, \citenamefont
  {Kirejczyk}, \citenamefont {Korolija}, \citenamefont {Kotte}, \citenamefont
  {Lebedev}, \citenamefont {Lopez}, \citenamefont {Matulewicz}, \citenamefont
  {Neubert}, \citenamefont {Petrovici}, \citenamefont {Rami}, \citenamefont
  {Ryu}, \citenamefont {Seres}, \citenamefont {Sikora}, \citenamefont {Sim},
  \citenamefont {Simion}, \citenamefont {Siwek-Wilczyńska}, \citenamefont
  {Smolyankin}, \citenamefont {Stoicea}, \citenamefont {Tymiński},
  \citenamefont {Wiśniewski}, \citenamefont {Wohlfarth}, \citenamefont {Xiao},
  \citenamefont {Xu}, \citenamefont {Yushmanov},\ and\ \citenamefont
  {Zhilin}}]{reisdorf_systematics_2010}%
  \BibitemOpen
  \bibfield  {author} {\bibinfo {author} {\bibfnamefont {W.}~\bibnamefont
  {Reisdorf}}, \bibinfo {author} {\bibfnamefont {A.}~\bibnamefont {Andronic}},
  \bibinfo {author} {\bibfnamefont {R.}~\bibnamefont {Averbeck}}, \bibinfo
  {author} {\bibfnamefont {M.}~\bibnamefont {Benabderrahmane}}, \bibinfo
  {author} {\bibfnamefont {O.}~\bibnamefont {Hartmann}}, \bibinfo {author}
  {\bibfnamefont {N.}~\bibnamefont {Herrmann}}, \bibinfo {author}
  {\bibfnamefont {K.}~\bibnamefont {Hildenbrand}}, \bibinfo {author}
  {\bibfnamefont {T.}~\bibnamefont {Kang}}, \bibinfo {author} {\bibfnamefont
  {Y.}~\bibnamefont {Kim}}, \bibinfo {author} {\bibfnamefont {M.}~\bibnamefont
  {Kiš}}, \bibinfo {author} {\bibfnamefont {P.}~\bibnamefont {Koczoń}},
  \bibinfo {author} {\bibfnamefont {T.}~\bibnamefont {Kress}}, \bibinfo
  {author} {\bibfnamefont {Y.}~\bibnamefont {Leifels}}, \bibinfo {author}
  {\bibfnamefont {M.}~\bibnamefont {Merschmeyer}}, \bibinfo {author}
  {\bibfnamefont {K.}~\bibnamefont {Piasecki}}, \bibinfo {author}
  {\bibfnamefont {A.}~\bibnamefont {Schüttauf}}, \bibinfo {author}
  {\bibfnamefont {M.}~\bibnamefont {Stockmeier}}, \bibinfo {author}
  {\bibfnamefont {V.}~\bibnamefont {Barret}}, \bibinfo {author} {\bibfnamefont
  {Z.}~\bibnamefont {Basrak}}, \bibinfo {author} {\bibfnamefont
  {N.}~\bibnamefont {Bastid}}, \bibinfo {author} {\bibfnamefont
  {R.}~\bibnamefont {Čaplar}}, \bibinfo {author} {\bibfnamefont
  {P.}~\bibnamefont {Crochet}}, \bibinfo {author} {\bibfnamefont
  {P.}~\bibnamefont {Dupieux}}, \bibinfo {author} {\bibfnamefont
  {M.}~\bibnamefont {Dželalija}}, \bibinfo {author} {\bibfnamefont
  {Z.}~\bibnamefont {Fodor}}, \bibinfo {author} {\bibfnamefont
  {P.}~\bibnamefont {Gasik}}, \bibinfo {author} {\bibfnamefont
  {Y.}~\bibnamefont {Grishkin}}, \bibinfo {author} {\bibfnamefont
  {B.}~\bibnamefont {Hong}}, \bibinfo {author} {\bibfnamefont {J.}~\bibnamefont
  {Kecskemeti}}, \bibinfo {author} {\bibfnamefont {M.}~\bibnamefont
  {Kirejczyk}}, \bibinfo {author} {\bibfnamefont {M.}~\bibnamefont {Korolija}},
  \bibinfo {author} {\bibfnamefont {R.}~\bibnamefont {Kotte}}, \bibinfo
  {author} {\bibfnamefont {A.}~\bibnamefont {Lebedev}}, \bibinfo {author}
  {\bibfnamefont {X.}~\bibnamefont {Lopez}}, \bibinfo {author} {\bibfnamefont
  {T.}~\bibnamefont {Matulewicz}}, \bibinfo {author} {\bibfnamefont
  {W.}~\bibnamefont {Neubert}}, \bibinfo {author} {\bibfnamefont
  {M.}~\bibnamefont {Petrovici}}, \bibinfo {author} {\bibfnamefont
  {F.}~\bibnamefont {Rami}}, \bibinfo {author} {\bibfnamefont {M.}~\bibnamefont
  {Ryu}}, \bibinfo {author} {\bibfnamefont {Z.}~\bibnamefont {Seres}}, \bibinfo
  {author} {\bibfnamefont {B.}~\bibnamefont {Sikora}}, \bibinfo {author}
  {\bibfnamefont {K.}~\bibnamefont {Sim}}, \bibinfo {author} {\bibfnamefont
  {V.}~\bibnamefont {Simion}}, \bibinfo {author} {\bibfnamefont
  {K.}~\bibnamefont {Siwek-Wilczyńska}}, \bibinfo {author} {\bibfnamefont
  {V.}~\bibnamefont {Smolyankin}}, \bibinfo {author} {\bibfnamefont
  {G.}~\bibnamefont {Stoicea}}, \bibinfo {author} {\bibfnamefont
  {Z.}~\bibnamefont {Tymiński}}, \bibinfo {author} {\bibfnamefont
  {K.}~\bibnamefont {Wiśniewski}}, \bibinfo {author} {\bibfnamefont
  {D.}~\bibnamefont {Wohlfarth}}, \bibinfo {author} {\bibfnamefont
  {Z.}~\bibnamefont {Xiao}}, \bibinfo {author} {\bibfnamefont {H.}~\bibnamefont
  {Xu}}, \bibinfo {author} {\bibfnamefont {I.}~\bibnamefont {Yushmanov}}, \
  and\ \bibinfo {author} {\bibfnamefont {A.}~\bibnamefont {Zhilin}},\ }\href
  {\doibase 10.1016/j.nuclphysa.2010.09.008} {\bibfield  {journal} {\bibinfo
  {journal} {Nuclear Physics A}\ }\textbf {\bibinfo {volume} {848}},\ \bibinfo
  {pages} {366} (\bibinfo {year} {2010}{\natexlab{a}})}\BibitemShut {NoStop}%
\bibitem [{\citenamefont {Danielewicz}(1995)}]{danielewicz_effects_1995}%
  \BibitemOpen
  \bibfield  {author} {\bibinfo {author} {\bibfnamefont {P.}~\bibnamefont
  {Danielewicz}},\ }\href {\doibase 10.1103/PhysRevC.51.716} {\bibfield
  {journal} {\bibinfo  {journal} {Phys. Rev. C}\ }\textbf {\bibinfo {volume}
  {51}},\ \bibinfo {pages} {716} (\bibinfo {year} {1995})}\BibitemShut
  {NoStop}%
\bibitem [{\citenamefont {Reisdorf}\ \emph
  {et~al.}(2010{\natexlab{b}})\citenamefont {Reisdorf}, \citenamefont
  {Andronic}, \citenamefont {Averbeck}, \citenamefont {Benabderrahmane},
  \citenamefont {Hartmann}, \citenamefont {Herrmann}, \citenamefont
  {Hildenbrand}, \citenamefont {Kang}, \citenamefont {Kim}, \citenamefont
  {Kiš}, \citenamefont {Koczoń}, \citenamefont {Kress}, \citenamefont
  {Leifels}, \citenamefont {Merschmeyer}, \citenamefont {Piasecki},
  \citenamefont {Schüttauf}, \citenamefont {Stockmeier}, \citenamefont
  {Barret}, \citenamefont {Basrak}, \citenamefont {Bastid}, \citenamefont
  {Čaplar}, \citenamefont {Crochet}, \citenamefont {Dupieux}, \citenamefont
  {Dželalija}, \citenamefont {Fodor}, \citenamefont {Gasik}, \citenamefont
  {Grishkin}, \citenamefont {Hong}, \citenamefont {Kecskemeti}, \citenamefont
  {Kirejczyk}, \citenamefont {Korolija}, \citenamefont {Kotte}, \citenamefont
  {Lebedev}, \citenamefont {Lopez}, \citenamefont {Matulewicz}, \citenamefont
  {Neubert}, \citenamefont {Petrovici}, \citenamefont {Rami}, \citenamefont
  {Ryu}, \citenamefont {Seres}, \citenamefont {Sikora}, \citenamefont {Sim},
  \citenamefont {Simion}, \citenamefont {Siwek-Wilczyńska}, \citenamefont
  {Smolyankin}, \citenamefont {Stoicea}, \citenamefont {Tymiński},
  \citenamefont {Wiśniewski}, \citenamefont {Wohlfarth}, \citenamefont {Xiao},
  \citenamefont {Xu}, \citenamefont {Yushmanov},\ and\ \citenamefont
  {Zhilin}}]{my_reisdorf_systematics_2010}%
  \BibitemOpen
  \bibfield  {author} {\bibinfo {author} {\bibfnamefont {W.}~\bibnamefont
  {Reisdorf}}, \bibinfo {author} {\bibfnamefont {A.}~\bibnamefont {Andronic}},
  \bibinfo {author} {\bibfnamefont {R.}~\bibnamefont {Averbeck}}, \bibinfo
  {author} {\bibfnamefont {M.}~\bibnamefont {Benabderrahmane}}, \bibinfo
  {author} {\bibfnamefont {O.}~\bibnamefont {Hartmann}}, \bibinfo {author}
  {\bibfnamefont {N.}~\bibnamefont {Herrmann}}, \bibinfo {author}
  {\bibfnamefont {K.}~\bibnamefont {Hildenbrand}}, \bibinfo {author}
  {\bibfnamefont {T.}~\bibnamefont {Kang}}, \bibinfo {author} {\bibfnamefont
  {Y.}~\bibnamefont {Kim}}, \bibinfo {author} {\bibfnamefont {M.}~\bibnamefont
  {Kiš}}, \bibinfo {author} {\bibfnamefont {P.}~\bibnamefont {Koczoń}},
  \bibinfo {author} {\bibfnamefont {T.}~\bibnamefont {Kress}}, \bibinfo
  {author} {\bibfnamefont {Y.}~\bibnamefont {Leifels}}, \bibinfo {author}
  {\bibfnamefont {M.}~\bibnamefont {Merschmeyer}}, \bibinfo {author}
  {\bibfnamefont {K.}~\bibnamefont {Piasecki}}, \bibinfo {author}
  {\bibfnamefont {A.}~\bibnamefont {Schüttauf}}, \bibinfo {author}
  {\bibfnamefont {M.}~\bibnamefont {Stockmeier}}, \bibinfo {author}
  {\bibfnamefont {V.}~\bibnamefont {Barret}}, \bibinfo {author} {\bibfnamefont
  {Z.}~\bibnamefont {Basrak}}, \bibinfo {author} {\bibfnamefont
  {N.}~\bibnamefont {Bastid}}, \bibinfo {author} {\bibfnamefont
  {R.}~\bibnamefont {Čaplar}}, \bibinfo {author} {\bibfnamefont
  {P.}~\bibnamefont {Crochet}}, \bibinfo {author} {\bibfnamefont
  {P.}~\bibnamefont {Dupieux}}, \bibinfo {author} {\bibfnamefont
  {M.}~\bibnamefont {Dželalija}}, \bibinfo {author} {\bibfnamefont
  {Z.}~\bibnamefont {Fodor}}, \bibinfo {author} {\bibfnamefont
  {P.}~\bibnamefont {Gasik}}, \bibinfo {author} {\bibfnamefont
  {Y.}~\bibnamefont {Grishkin}}, \bibinfo {author} {\bibfnamefont
  {B.}~\bibnamefont {Hong}}, \bibinfo {author} {\bibfnamefont {J.}~\bibnamefont
  {Kecskemeti}}, \bibinfo {author} {\bibfnamefont {M.}~\bibnamefont
  {Kirejczyk}}, \bibinfo {author} {\bibfnamefont {M.}~\bibnamefont {Korolija}},
  \bibinfo {author} {\bibfnamefont {R.}~\bibnamefont {Kotte}}, \bibinfo
  {author} {\bibfnamefont {A.}~\bibnamefont {Lebedev}}, \bibinfo {author}
  {\bibfnamefont {X.}~\bibnamefont {Lopez}}, \bibinfo {author} {\bibfnamefont
  {T.}~\bibnamefont {Matulewicz}}, \bibinfo {author} {\bibfnamefont
  {W.}~\bibnamefont {Neubert}}, \bibinfo {author} {\bibfnamefont
  {M.}~\bibnamefont {Petrovici}}, \bibinfo {author} {\bibfnamefont
  {F.}~\bibnamefont {Rami}}, \bibinfo {author} {\bibfnamefont {M.}~\bibnamefont
  {Ryu}}, \bibinfo {author} {\bibfnamefont {Z.}~\bibnamefont {Seres}}, \bibinfo
  {author} {\bibfnamefont {B.}~\bibnamefont {Sikora}}, \bibinfo {author}
  {\bibfnamefont {K.}~\bibnamefont {Sim}}, \bibinfo {author} {\bibfnamefont
  {V.}~\bibnamefont {Simion}}, \bibinfo {author} {\bibfnamefont
  {K.}~\bibnamefont {Siwek-Wilczyńska}}, \bibinfo {author} {\bibfnamefont
  {V.}~\bibnamefont {Smolyankin}}, \bibinfo {author} {\bibfnamefont
  {G.}~\bibnamefont {Stoicea}}, \bibinfo {author} {\bibfnamefont
  {Z.}~\bibnamefont {Tymiński}}, \bibinfo {author} {\bibfnamefont
  {K.}~\bibnamefont {Wiśniewski}}, \bibinfo {author} {\bibfnamefont
  {D.}~\bibnamefont {Wohlfarth}}, \bibinfo {author} {\bibfnamefont
  {Z.}~\bibnamefont {Xiao}}, \bibinfo {author} {\bibfnamefont {H.}~\bibnamefont
  {Xu}}, \bibinfo {author} {\bibfnamefont {I.}~\bibnamefont {Yushmanov}}, \
  and\ \bibinfo {author} {\bibfnamefont {A.}~\bibnamefont {Zhilin}},\ }\href
  {\doibase 10.1016/j.nuclphysa.2010.09.008} {\bibfield  {journal} {\bibinfo
  {journal} {Nuclear Physics A}\ }\textbf {\bibinfo {volume} {848}},\ \bibinfo
  {pages} {366} (\bibinfo {year} {2010}{\natexlab{b}})}\BibitemShut {NoStop}%
\bibitem [{\citenamefont {Rami}\ \emph {et~al.}(2000)\citenamefont {Rami},
  \citenamefont {Leifels}, \citenamefont {de~Schauenburg}, \citenamefont
  {Gobbi}, \citenamefont {Hong}, \citenamefont {Alard}, \citenamefont
  {Andronic}, \citenamefont {Averbeck}, \citenamefont {Barret}, \citenamefont
  {Basrak}, \citenamefont {Bastid}, \citenamefont {Belyaev}, \citenamefont
  {Bendarag}, \citenamefont {Berek}, \citenamefont {Čaplar}, \citenamefont
  {Cindro}, \citenamefont {Crochet}, \citenamefont {Devismes}, \citenamefont
  {Dupieux}, \citenamefont {Dželalija}, \citenamefont {Eskef}, \citenamefont
  {Finck}, \citenamefont {Fodor}, \citenamefont {Folger}, \citenamefont
  {Fraysse}, \citenamefont {Genoux-Lubain}, \citenamefont {Grigorian},
  \citenamefont {Grishkin}, \citenamefont {Herrmann}, \citenamefont
  {Hildenbrand}, \citenamefont {Kecskemeti}, \citenamefont {Kim}, \citenamefont
  {Koczon}, \citenamefont {Kirejczyk}, \citenamefont {Korolija}, \citenamefont
  {Kotte}, \citenamefont {Kowalczyk}, \citenamefont {Kress}, \citenamefont
  {Kutsche}, \citenamefont {Lebedev}, \citenamefont {Lee}, \citenamefont
  {Manko}, \citenamefont {Merlitz}, \citenamefont {Mohren}, \citenamefont
  {Moisa}, \citenamefont {Mösner}, \citenamefont {Neubert}, \citenamefont
  {Nianine}, \citenamefont {Pelte}, \citenamefont {Petrovici}, \citenamefont
  {Pinkenburg}, \citenamefont {Plettner}, \citenamefont {Reisdorf},
  \citenamefont {Ritman}, \citenamefont {Schüll}, \citenamefont {Seres},
  \citenamefont {Sikora}, \citenamefont {Sim}, \citenamefont {Simion},
  \citenamefont {Siwek-Wilczyńska}, \citenamefont {Somov}, \citenamefont
  {Stockmeier}, \citenamefont {Stoicea}, \citenamefont {Vasiliev},
  \citenamefont {Wagner}, \citenamefont {Wiśniewski}, \citenamefont
  {Wohlfarth}, \citenamefont {Yang}, \citenamefont {Yushmanov}, \citenamefont
  {Zhilin},\ and\ \citenamefont {Collaboration)}}]{rami_isospin_2000}%
  \BibitemOpen
  \bibfield  {author} {\bibinfo {author} {\bibfnamefont {F.}~\bibnamefont
  {Rami}}, \bibinfo {author} {\bibfnamefont {Y.}~\bibnamefont {Leifels}},
  \bibinfo {author} {\bibfnamefont {B.}~\bibnamefont {de~Schauenburg}},
  \bibinfo {author} {\bibfnamefont {A.}~\bibnamefont {Gobbi}}, \bibinfo
  {author} {\bibfnamefont {B.}~\bibnamefont {Hong}}, \bibinfo {author}
  {\bibfnamefont {J.~P.}\ \bibnamefont {Alard}}, \bibinfo {author}
  {\bibfnamefont {A.}~\bibnamefont {Andronic}}, \bibinfo {author}
  {\bibfnamefont {R.}~\bibnamefont {Averbeck}}, \bibinfo {author}
  {\bibfnamefont {V.}~\bibnamefont {Barret}}, \bibinfo {author} {\bibfnamefont
  {Z.}~\bibnamefont {Basrak}}, \bibinfo {author} {\bibfnamefont
  {N.}~\bibnamefont {Bastid}}, \bibinfo {author} {\bibfnamefont
  {I.}~\bibnamefont {Belyaev}}, \bibinfo {author} {\bibfnamefont
  {A.}~\bibnamefont {Bendarag}}, \bibinfo {author} {\bibfnamefont
  {G.}~\bibnamefont {Berek}}, \bibinfo {author} {\bibfnamefont
  {R.}~\bibnamefont {Čaplar}}, \bibinfo {author} {\bibfnamefont
  {N.}~\bibnamefont {Cindro}}, \bibinfo {author} {\bibfnamefont
  {P.}~\bibnamefont {Crochet}}, \bibinfo {author} {\bibfnamefont
  {A.}~\bibnamefont {Devismes}}, \bibinfo {author} {\bibfnamefont
  {P.}~\bibnamefont {Dupieux}}, \bibinfo {author} {\bibfnamefont
  {M.}~\bibnamefont {Dželalija}}, \bibinfo {author} {\bibfnamefont
  {M.}~\bibnamefont {Eskef}}, \bibinfo {author} {\bibfnamefont
  {C.}~\bibnamefont {Finck}}, \bibinfo {author} {\bibfnamefont
  {Z.}~\bibnamefont {Fodor}}, \bibinfo {author} {\bibfnamefont
  {H.}~\bibnamefont {Folger}}, \bibinfo {author} {\bibfnamefont
  {L.}~\bibnamefont {Fraysse}}, \bibinfo {author} {\bibfnamefont
  {A.}~\bibnamefont {Genoux-Lubain}}, \bibinfo {author} {\bibfnamefont
  {Y.}~\bibnamefont {Grigorian}}, \bibinfo {author} {\bibfnamefont
  {Y.}~\bibnamefont {Grishkin}}, \bibinfo {author} {\bibfnamefont
  {N.}~\bibnamefont {Herrmann}}, \bibinfo {author} {\bibfnamefont {K.~D.}\
  \bibnamefont {Hildenbrand}}, \bibinfo {author} {\bibfnamefont
  {J.}~\bibnamefont {Kecskemeti}}, \bibinfo {author} {\bibfnamefont {Y.~J.}\
  \bibnamefont {Kim}}, \bibinfo {author} {\bibfnamefont {P.}~\bibnamefont
  {Koczon}}, \bibinfo {author} {\bibfnamefont {M.}~\bibnamefont {Kirejczyk}},
  \bibinfo {author} {\bibfnamefont {M.}~\bibnamefont {Korolija}}, \bibinfo
  {author} {\bibfnamefont {R.}~\bibnamefont {Kotte}}, \bibinfo {author}
  {\bibfnamefont {M.}~\bibnamefont {Kowalczyk}}, \bibinfo {author}
  {\bibfnamefont {T.}~\bibnamefont {Kress}}, \bibinfo {author} {\bibfnamefont
  {R.}~\bibnamefont {Kutsche}}, \bibinfo {author} {\bibfnamefont
  {A.}~\bibnamefont {Lebedev}}, \bibinfo {author} {\bibfnamefont {K.~S.}\
  \bibnamefont {Lee}}, \bibinfo {author} {\bibfnamefont {V.}~\bibnamefont
  {Manko}}, \bibinfo {author} {\bibfnamefont {H.}~\bibnamefont {Merlitz}},
  \bibinfo {author} {\bibfnamefont {S.}~\bibnamefont {Mohren}}, \bibinfo
  {author} {\bibfnamefont {D.}~\bibnamefont {Moisa}}, \bibinfo {author}
  {\bibfnamefont {J.}~\bibnamefont {Mösner}}, \bibinfo {author} {\bibfnamefont
  {W.}~\bibnamefont {Neubert}}, \bibinfo {author} {\bibfnamefont
  {A.}~\bibnamefont {Nianine}}, \bibinfo {author} {\bibfnamefont
  {D.}~\bibnamefont {Pelte}}, \bibinfo {author} {\bibfnamefont
  {M.}~\bibnamefont {Petrovici}}, \bibinfo {author} {\bibfnamefont
  {C.}~\bibnamefont {Pinkenburg}}, \bibinfo {author} {\bibfnamefont
  {C.}~\bibnamefont {Plettner}}, \bibinfo {author} {\bibfnamefont
  {W.}~\bibnamefont {Reisdorf}}, \bibinfo {author} {\bibfnamefont
  {J.}~\bibnamefont {Ritman}}, \bibinfo {author} {\bibfnamefont
  {D.}~\bibnamefont {Schüll}}, \bibinfo {author} {\bibfnamefont
  {Z.}~\bibnamefont {Seres}}, \bibinfo {author} {\bibfnamefont
  {B.}~\bibnamefont {Sikora}}, \bibinfo {author} {\bibfnamefont {K.~S.}\
  \bibnamefont {Sim}}, \bibinfo {author} {\bibfnamefont {V.}~\bibnamefont
  {Simion}}, \bibinfo {author} {\bibfnamefont {K.}~\bibnamefont
  {Siwek-Wilczyńska}}, \bibinfo {author} {\bibfnamefont {A.}~\bibnamefont
  {Somov}}, \bibinfo {author} {\bibfnamefont {M.~R.}\ \bibnamefont
  {Stockmeier}}, \bibinfo {author} {\bibfnamefont {G.}~\bibnamefont {Stoicea}},
  \bibinfo {author} {\bibfnamefont {M.}~\bibnamefont {Vasiliev}}, \bibinfo
  {author} {\bibfnamefont {P.}~\bibnamefont {Wagner}}, \bibinfo {author}
  {\bibfnamefont {K.}~\bibnamefont {Wiśniewski}}, \bibinfo {author}
  {\bibfnamefont {D.}~\bibnamefont {Wohlfarth}}, \bibinfo {author}
  {\bibfnamefont {J.~T.}\ \bibnamefont {Yang}}, \bibinfo {author}
  {\bibfnamefont {I.}~\bibnamefont {Yushmanov}}, \bibinfo {author}
  {\bibfnamefont {A.}~\bibnamefont {Zhilin}}, \ and\ \bibinfo {author}
  {\bibfnamefont {F.}~\bibnamefont {Collaboration)}},\ }\href {\doibase
  10.1103/PhysRevLett.84.1120} {\bibfield  {journal} {\bibinfo  {journal}
  {Phys. Rev. Lett.}\ }\textbf {\bibinfo {volume} {84}},\ \bibinfo {pages}
  {1120} (\bibinfo {year} {2000})}\BibitemShut {NoStop}%
\bibitem [{\citenamefont {Blaizot}\ \emph {et~al.}(1995)\citenamefont
  {Blaizot}, \citenamefont {Berger}, \citenamefont {Dechargé},\ and\
  \citenamefont {Girod}}]{blaizot_microscopic_1995}%
  \BibitemOpen
  \bibfield  {author} {\bibinfo {author} {\bibfnamefont {J.}~\bibnamefont
  {Blaizot}}, \bibinfo {author} {\bibfnamefont {J.}~\bibnamefont {Berger}},
  \bibinfo {author} {\bibfnamefont {J.}~\bibnamefont {Dechargé}}, \ and\
  \bibinfo {author} {\bibfnamefont {M.}~\bibnamefont {Girod}},\ }\href
  {\doibase 10.1016/0375-9474(95)00294-B} {\bibfield  {journal} {\bibinfo
  {journal} {Nuclear Physics A}\ }\textbf {\bibinfo {volume} {591}},\ \bibinfo
  {pages} {435} (\bibinfo {year} {1995})}\BibitemShut {NoStop}%
\bibitem [{\citenamefont {Hartnack}\ \emph {et~al.}(1998)\citenamefont
  {Hartnack}, \citenamefont {Puri}, \citenamefont {Aichelin}, \citenamefont
  {Konopka}, \citenamefont {Bass}, \citenamefont {Stöcker},\ and\
  \citenamefont {Greiner}}]{hartnack_modelling_1998}%
  \BibitemOpen
  \bibfield  {author} {\bibinfo {author} {\bibfnamefont {C.}~\bibnamefont
  {Hartnack}}, \bibinfo {author} {\bibfnamefont {R.~K.}\ \bibnamefont {Puri}},
  \bibinfo {author} {\bibfnamefont {J.}~\bibnamefont {Aichelin}}, \bibinfo
  {author} {\bibfnamefont {J.}~\bibnamefont {Konopka}}, \bibinfo {author}
  {\bibfnamefont {S.~A.}\ \bibnamefont {Bass}}, \bibinfo {author}
  {\bibfnamefont {H.}~\bibnamefont {Stöcker}}, \ and\ \bibinfo {author}
  {\bibfnamefont {W.}~\bibnamefont {Greiner}},\ }\href {\doibase
  10.1007/s100500050045} {\bibfield  {journal} {\bibinfo  {journal} {EPJ A}\
  }\textbf {\bibinfo {volume} {1}},\ \bibinfo {pages} {151} (\bibinfo {year}
  {1998})}\BibitemShut {NoStop}%
\bibitem [{\citenamefont {Danielewicz}\ \emph {et~al.}(2002)\citenamefont
  {Danielewicz}, \citenamefont {Lacey},\ and\ \citenamefont
  {Lynch}}]{danielewicz_determination_2002}%
  \BibitemOpen
  \bibfield  {author} {\bibinfo {author} {\bibfnamefont {P.}~\bibnamefont
  {Danielewicz}}, \bibinfo {author} {\bibfnamefont {R.}~\bibnamefont {Lacey}},
  \ and\ \bibinfo {author} {\bibfnamefont {W.~G.}\ \bibnamefont {Lynch}},\
  }\href {\doibase 10.1126/science.1078070} {\bibfield  {journal} {\bibinfo
  {journal} {Science}\ }\textbf {\bibinfo {volume} {298}},\ \bibinfo {pages}
  {1592} (\bibinfo {year} {2002})}\BibitemShut {NoStop}%
\bibitem [{\citenamefont {Bertsch}\ and\ \citenamefont
  {Cugnon}(1981)}]{bertsch_entropy_1981}%
  \BibitemOpen
  \bibfield  {author} {\bibinfo {author} {\bibfnamefont {G.}~\bibnamefont
  {Bertsch}}\ and\ \bibinfo {author} {\bibfnamefont {J.}~\bibnamefont
  {Cugnon}},\ }\href {\doibase 10.1103/PhysRevC.24.2514} {\bibfield  {journal}
  {\bibinfo  {journal} {Phys. Rev. C}\ }\textbf {\bibinfo {volume} {24}},\
  \bibinfo {pages} {2514} (\bibinfo {year} {1981})}\BibitemShut {NoStop}%
\bibitem [{\citenamefont {Csernai}\ and\ \citenamefont
  {Kapusta}(1986)}]{csernai_entropy_1986}%
  \BibitemOpen
  \bibfield  {author} {\bibinfo {author} {\bibfnamefont {L.}~\bibnamefont
  {Csernai}}\ and\ \bibinfo {author} {\bibfnamefont {J.~I.}\ \bibnamefont
  {Kapusta}},\ }\href {\doibase 10.1016/0370-1573(86)90031-1} {\bibfield
  {journal} {\bibinfo  {journal} {Physics Reports}\ }\textbf {\bibinfo {volume}
  {131}},\ \bibinfo {pages} {223} (\bibinfo {year} {1986})}\BibitemShut
  {NoStop}%
\bibitem [{\citenamefont {Bertsch}(1983)}]{bertsch_entropy_1983}%
  \BibitemOpen
  \bibfield  {author} {\bibinfo {author} {\bibfnamefont {G.~F.}\ \bibnamefont
  {Bertsch}},\ }\href {\doibase 10.1016/0375-9474(83)90435-9} {\bibfield
  {journal} {\bibinfo  {journal} {Nuclear Physics A}\ }\textbf {\bibinfo
  {volume} {400}},\ \bibinfo {pages} {221} (\bibinfo {year}
  {1983})}\BibitemShut {NoStop}%
\bibitem [{\citenamefont {Bernhard}\ \emph {et~al.}(2015)\citenamefont
  {Bernhard}, \citenamefont {Marcy}, \citenamefont {Coleman-Smith},
  \citenamefont {Huzurbazar}, \citenamefont {Wolpert},\ and\ \citenamefont
  {Bass}}]{bernhard_quantifying_2015}%
  \BibitemOpen
  \bibfield  {author} {\bibinfo {author} {\bibfnamefont {J.~E.}\ \bibnamefont
  {Bernhard}}, \bibinfo {author} {\bibfnamefont {P.~W.}\ \bibnamefont {Marcy}},
  \bibinfo {author} {\bibfnamefont {C.~E.}\ \bibnamefont {Coleman-Smith}},
  \bibinfo {author} {\bibfnamefont {S.}~\bibnamefont {Huzurbazar}}, \bibinfo
  {author} {\bibfnamefont {R.~L.}\ \bibnamefont {Wolpert}}, \ and\ \bibinfo
  {author} {\bibfnamefont {S.~A.}\ \bibnamefont {Bass}},\ }\href {\doibase
  10.1103/PhysRevC.91.054910} {\bibfield  {journal} {\bibinfo  {journal} {Phys.
  Rev. C}\ }\textbf {\bibinfo {volume} {91}},\ \bibinfo {pages} {054910}
  (\bibinfo {year} {2015})}\BibitemShut {NoStop}%
\bibitem [{\citenamefont {Miller}\ \emph {et~al.}(2007)\citenamefont {Miller},
  \citenamefont {Reygers}, \citenamefont {Sanders},\ and\ \citenamefont
  {Steinberg}}]{miller_glauber_2007}%
  \BibitemOpen
  \bibfield  {author} {\bibinfo {author} {\bibfnamefont {M.~L.}\ \bibnamefont
  {Miller}}, \bibinfo {author} {\bibfnamefont {K.}~\bibnamefont {Reygers}},
  \bibinfo {author} {\bibfnamefont {S.~J.}\ \bibnamefont {Sanders}}, \ and\
  \bibinfo {author} {\bibfnamefont {P.}~\bibnamefont {Steinberg}},\ }\href
  {\doibase 10.1146/annurev.nucl.57.090506.123020} {\bibfield  {journal}
  {\bibinfo  {journal} {Annu. Rev. Nucl. Part. Sci.}\ }\textbf {\bibinfo
  {volume} {57}},\ \bibinfo {pages} {205} (\bibinfo {year} {2007})}\BibitemShut
  {NoStop}%
\bibitem [{\citenamefont {Adil}\ \emph {et~al.}(2006)\citenamefont {Adil},
  \citenamefont {Drescher}, \citenamefont {Dumitru}, \citenamefont
  {Hayashigaki},\ and\ \citenamefont {Nara}}]{adil_eccentricity_2006}%
  \BibitemOpen
  \bibfield  {author} {\bibinfo {author} {\bibfnamefont {A.}~\bibnamefont
  {Adil}}, \bibinfo {author} {\bibfnamefont {H.-J.}\ \bibnamefont {Drescher}},
  \bibinfo {author} {\bibfnamefont {A.}~\bibnamefont {Dumitru}}, \bibinfo
  {author} {\bibfnamefont {A.}~\bibnamefont {Hayashigaki}}, \ and\ \bibinfo
  {author} {\bibfnamefont {Y.}~\bibnamefont {Nara}},\ }\href {\doibase
  10.1103/PhysRevC.74.044905} {\bibfield  {journal} {\bibinfo  {journal} {Phys.
  Rev. C}\ }\textbf {\bibinfo {volume} {74}},\ \bibinfo {pages} {044905}
  (\bibinfo {year} {2006})}\BibitemShut {NoStop}%
\bibitem [{\citenamefont {Karsch}\ \emph {et~al.}(2001)\citenamefont {Karsch},
  \citenamefont {Laermann},\ and\ \citenamefont {Peikert}}]{karsch_quark_2001}%
  \BibitemOpen
  \bibfield  {author} {\bibinfo {author} {\bibfnamefont {F.}~\bibnamefont
  {Karsch}}, \bibinfo {author} {\bibfnamefont {E.}~\bibnamefont {Laermann}}, \
  and\ \bibinfo {author} {\bibfnamefont {A.}~\bibnamefont {Peikert}},\ }\href
  {\doibase 10.1016/S0550-3213(01)00200-0} {\bibfield  {journal} {\bibinfo
  {journal} {Nuclear Physics B}\ }\textbf {\bibinfo {volume} {605}},\ \bibinfo
  {pages} {579} (\bibinfo {year} {2001})}\BibitemShut {NoStop}%
\end{thebibliography}%

\end{document}